\documentclass[aps,twocolumn,preprintnumbers,amsmath,amssymb,floatfix, superscriptaddress]{revtex4}

\usepackage{graphicx}
\usepackage{dcolumn}

\usepackage{bm}
\begin{document}

\title{Multiple glass transitions  in star polymer mixtures: \\
Insights from theory and simulations}

\author{Christian Mayer}
\affiliation{Dipartimento di Fisica and CNR-INFM-SOFT, Universit{\`a} di Roma La Sapienza, Piazzale Aldo Moro 2, I-00185 Rome, Italy}
\affiliation{Institut f{\"u}r Theoretische Physik II: Weiche Materie,
             Heinrich-Heine-Universit{\"a}t D{\"u}sseldorf,
             Universit{\"a}tsstra{\ss}e 1, D-40225 D{\"u}sseldorf,
             Germany}

\author{Francesco Sciortino} 
\affiliation{Dipartimento di Fisica and CNR-INFM-SOFT, Universit{\`a} di Roma La Sapienza, Piazzale Aldo Moro 2, I-00185 Rome, Italy}

\author{Christos~N.~Likos}
\affiliation{Institut f{\"u}r Theoretische Physik II: Weiche Materie,
             Heinrich-Heine-Universit{\"a}t D{\"u}sseldorf,
             Universit{\"a}tsstra{\ss}e 1, D-40225 D{\"u}sseldorf,
             Germany}
\affiliation{The Erwin Schr{\"o}dinger International Institute
for Mathematical Physics (ESI), Boltzmanngasse 9, A-1090 Vienna, Austria}
\affiliation{Institut f{\"u}r Theoretische Physik, Technische 
Universit{\"a}t Wien, Wiedner Hauptstra{\ss}e 8-10, A-1040 Wien, Austria}

\author{Piero Tartaglia}
\affiliation{Dipartimento di Fisica and CNR-INFM-SMC, Universit{\`a} di Roma La Sapienza, Piazzale Aldo Moro 2, I-00185 Rome, Italy}     

\author{Hartmut~L\"owen}
\affiliation{Institut f{\"u}r Theoretische Physik II: Weiche Materie,
             Heinrich-Heine-Universit{\"a}t D{\"u}sseldorf,
             Universit{\"a}tsstra{\ss}e 1, D-40225 D{\"u}sseldorf,
             Germany}

\author{Emanuela Zaccarelli}
\affiliation{Dipartimento di Fisica and CNR-INFM-SOFT, Universit{\`a} di Roma La Sapienza, Piazzale Aldo Moro 2, I-00185 Rome, Italy}

\date{\today} 

\begin{abstract}

The glass transition in binary mixtures of star polymers is studied by
mode coupling theory and extensive molecular dynamics computer
simulations.  In particular, we have explored vitrification in the
parameter space of size asymmetry $\delta$ and concentration $\rho_2$
of the small star polymers at fixed concentration of the large
ones. Depending on the choice of parameters, three different glassy
states are identified: a single glass of big polymers at low $\delta$
and low $\rho_2$, a double glass at high $\delta$ and low $\rho_2$,
and a novel double glass at high $\rho_2$ and high $\delta$ which is
characterized by a strong localization of the small particles.  At low
$\delta$ and high $\rho_2$ there is a competition between
vitrification and phase separation.  Centered in the $(\delta,
\rho_2)$-plane, a liquid lake shows up revealing reentrant glass
formation.  We compare the behavior of the dynamical density
correlators with the predictions of the theory and find remarkable
agreement between the two.
\end{abstract}

\maketitle

\section{Introduction}
The slowing down of the dynamics in a supercooled liquid near the
kinetic glass transition still requires a complete molecular
understanding though recent progress has been made by
experiments\cite{Pusey,Cipeletti, scienceberthier}, by computer
simulations
\cite{Klein,Binder,Baschnagel,sciortino_review,zacca_review,reviewheuerjpcm}
and theory \cite{goetze91,Gotze2,chandler,entropydroplet}.  The
inherent polydispersity of colloidal systems facilitates the formation
of disordered arrested states, as compared to the formation of the
underlying equilibrium crystal \cite{Pusey}.  In simulations, it is
difficult to vitrify a pure one-component system of spherical
particles due to the onset of crystallization, so that binary mixtures
constitute a much richer and more convenient system to study the glass
transition \cite{kobandersen}. The thermodynamically stable state of
an homogeneous undercooled binary melt typically involves two
phase-separated binary crystals, A-rich and B-rich
\cite{HenleyLikos,Fernandez} or a complex alloy in which the two
components are mixed \cite{HenleyLikos,Julia}. In either case, the
processes involved to reach true thermodynamic equilibrium require
collective, mutually coordinated motions that take enormously long
time. Therefore, the glass transition does take place also in
simulation, since crystallization is prevented by the inherent
polydispersity of the mixture. Hence, either binary or polydisperse
systems have always been used as prototype models for the glass
transition, for example repulsive soft spheres
\cite{Roux,Klein,barrat90} or Lennard-Jones mixtures
\cite{Ngai,Vollmayr,Harrowell}, charged mixtures \cite{Bosse}, hard
spheres \cite{Heuer}, and Yukawa systems \cite{Lowen, sciortino05}.
In fact, the main reason for considering these {\it slightly
  asymmetric} binary mixtures is that crystallization was avoided.
Upon cooling down typically all species freeze into a glass at the
same temperature.

More recently, insight has been gained into the glass transition 
of {\it strongly asymmetric} mixtures consisting
of large colloidal spheres and small additives such as 
smaller colloids or non-adsorbing polymers which induce an effective depletion attraction between
the large particles whose range is dictated by the size of the small 
particles. For strong
asymmetries, very short range attractions (relative to the core size of 
the large particles)
can thus be realized. A new type of glass was observed for these attractive 
interactions --- termed an {\it attractive} glass  --- in 
experiments of colloid-polymer mixtures \cite{pham} and in computer simulations 
\cite{Zac01PRE}, following predictions based on  mode coupling theory \cite{ber99a}.
 There is an interesting reentrant behavior \cite{sciortinonatmat}
 for increasing concentration of the additives:
the {\it repulsive glass} (at vanishing additive concentration, i.e.\
 in the absence of any attraction)  melts upon a threshold additive 
concentration.
The resulting melt revitrifies  into a second type of glass, namely the 
{\it attractive glass}, for higher
additive concentrations.
Both for repulsive and  attractive glass, only the large particles 
freeze into a disordered glassy matrix
while the small additives always remain mobile and stay in a fluid state
occupying the ``holes'' within the glassy matrix. 
This has been made explicit by computer simulations and mode coupling 
theory for the Asakura-Oosawa model in which the polymer-polymer interaction
is assumed to be ideal.
It was found \cite{Zacca_PRL} that the bare mobility of the additives is very 
important:
when it is comparable to that of the colloids there is a freezing
into a common glassy matrix of colloids and additives, while  when it is 
much larger than the one of the colloids, freezing involves only the colloid component.

Experiments on binary hard sphere mixtures \cite{imhof95} have shown
that there is a geometrically-inspired cross-over behavior in the
dynamics of the small spheres when the ratio $\delta$ of the two
spheres diameter is about 0.15. This is well below the size asymmetry
considered in earlier mode-coupling calculations
\cite{Voigtmann}. Another system in which dynamical arrest has been
studied theoretically and experimentally are star polymers and
micellar solutions \cite{foffi03,loppinet01,laurati05,laurati07}.  The
first systematic attempt to vary the size asymmetry and to monitor its
influence on the glass transition was performed in a binary mixture of
star polymers in a good solvent \cite{Zacca2005}. Mode coupling theory
and experimental data were compared for different size ratios $\delta$
of the star polymer species.  At low $\delta$ the small additives
remain ergodic in the glassy matrix formed by the large stars
(\textit{single glass}) while at high $\delta$ both species become
arrested simultaneously (\textit{double glass}).

For increasing concentration of the smaller star polymers a melting
into a fluid was found.  The threshold concentration of the smaller
stars for melting behaves non-monotonically as a function of $\delta$
for fixed concentration of the big stars with a $U$-type shape.  This
non-monotonicity is a result of the ultrasoft effective interactions
between the star polymers \cite{likos} and has no counterpart in
hard-sphere based systems.  Additive induced melting has also been
found in mixtures of star polymers and liner polymer chains
\cite{manolis:prl:02, mayer07a}.  More recently, a detailed study of
the structural properties at high densities of the additives was
performed \cite{unpublished} and an additional double glass, called
\textit{asymmetric glass}, has been identified.

In the present paper we elaborate more on the system of binary star
polymers. Our motivation to do so is twofold: Firstly, aiming at a
deeper understanding of the full dynamical behavior of collective- and
self-correlations close to the various glass transitions, we performed
extensive molecular dynamics computer simulations and compared the
results with predictions of Mode Coupling Theory calculations. In this
fashion, we were able to extract the dynamical scaling laws by fitting
the simulation data.  As regards the arrested plateau values of the
dynamical correlations and localization lengths of both large and
small components we find quantitative agreement between simulations
and MCT.  Secondly, we have expanded the parameter space up to
concentrations of added small stars higher than previously studied
\cite{Zacca2005} and found a reentrant vitrification of the molten
state \cite{unpublished}.  In the plane of size asymmetry $\delta$ and
added concentration $\rho_2$ the molten state formes a lake with an
$O$-shape.  In addition to the asymmetric glass, we also observe an
attractive glass at low $\delta$ and high $\rho_2$, which is unstable
with respect to phase separation.

The paper is organized as follows: in Sec.\ II
we introduce the theoretical methods used in this paper.
Sec.\ III contains the results based on Mode Coupling Theory, where we 
distinguish
four
regions in the $(\delta,\rho_2)$ plane. 
Simulation data for the glass transition 
line based on iso-diffusivity
curves are presented in Sec.\ IV,
whereas simulation data for the dynamical 
correlations
in the  four different regions are subsequently 
discussed in Sec.\ V. Finally, in Sec.\ VI we summarize
and present our conclusions.

\section{Theoretical Methods}

We describe the system in terms of effective interactions between the
star centers \cite{likos}. The monomer degrees of freedom have been
traced out in this approach.  This method is applicable to the problem
at hand, because we are interested in the arrest of the stars as whole
objects and not in the arrest of the monomers themeslves, which occurs
at much higher concentrations. Concomitantly, the monomers maintain
their fluctuating nature throughout. However, possible entanglements
of the arms are not included in this description.  The effective
interaction between star centers depends parametrically on the number
of arms (or functionality) $f$, and shows a logarithmic divergence
when the two stars overlap within their corona diameters $\sigma_i$,
$i = 1,2$,(where $\sigma \cong R_h$ \cite{manolis:prl:02}, with $R_h$
the hydrodynamic radius), followed by a Yukawa-tail for larger
separations \cite{likos}.

In this work we study binary star mixtures with
functionalities
$f_1=263$ and $f_2=64$. We express the mixture
composition by quoting
the values $\rho_1\sigma_1^3$ of the big and
$\rho_2\sigma_1^3$ of the
small stars, where $\rho_i$, $i=1,2$ are the respective number
densities.
The size ratio
$\delta=\sigma_2/\sigma_1$ is an additional 
parameter of the mixture.
The two-component generalization of the
aforementioned effective star-star potential reads
\cite{ferber},

\begin{equation}\label{eq:pot}
\begin{aligned}
\beta V_{ij}&=\Theta_{ij}\\ 
&\times\left\{
\begin{aligned}
&-\ln\left(\frac{r}{\sigma_{ij}}\right)+\frac{1}{1+\sigma_{ij}\kappa_{ij}}&\text{for
$r\leq\sigma_{ij}$;}\\
&\frac{1}{1+\sigma_{ij}\kappa_{ij}}\left(\frac{\sigma_{ij}}{r}\right)\exp(\sigma_{ij}\kappa_{ij}-r\kappa_{ij})
&\text{else,}
\end{aligned}
\right.
\end{aligned}
\end{equation}
where $\sigma_{ij}=(\sigma_i+\sigma_j)/2$,
$2/\kappa_{ij}=\sigma_i/\sqrt{f_i}+\sigma_j/\sqrt{f_j}$
and
\begin{equation}
\Theta_{ij}=\frac{5}{36}\frac{1}{\sqrt{2}-1}\left[(f_i+f_j)^{3/2}-(f_i^{3/2}+f_j^{3/2})\right].
\end{equation}

In order to calculate the fluid structure of the
mixture we solve the
binary Ornstein-Zernike (OZ) equation \cite{hansen06}
with the
Rogers-Young (RY) closure \cite{rogers84}.
This amounts to postulating the additional relation to hold
between the total correlation functions $h_{ij}(r)$ and their
direct counterparts $c_{ij}(r)$:
\begin{equation}
\begin{aligned}
h_{ij}(r)&=\exp\left[-\beta V_{ij}(r)\right]\\
&\times\left[1+\frac{\exp[(h_{ij}(r)-c_{ij}(r))f(r)]-1}{f(r)}\right] - 1,
\end{aligned}
\end{equation}
where $f(r)=1-\exp(-\alpha r)$ with $(\alpha > 0)$.  The
value of the
parameter $\alpha$ is determined by requiring the
equality of the
`fluctuation' and `virial' total compressibilities of
the system
\citep{rogers84,mayer04}.  For $\alpha\to 0$ the
Percus-Yevick closure
is obtained, while for $\alpha\to\infty$ the
hypernetted chain closure
(HNC) is recovered \cite{hansen06}.  The RY closure
has been shown to
correctly describe liquid properties for a number of
repulsive
potentials, including star polymers \cite{Watz},
spherical ramp \cite{Kumar05} and square
shoulder \cite{Lang} potentials.

From solving the OZ equation, it is also possible to
calculate the
partial static structure factors, describing the
density correlators
at equal time,
\begin{equation}
S_{ij}(q)=\delta_{ij}+\sqrt{\rho_i\rho_j}\tilde{h}_{ij}(q)=\frac{1}{\sqrt{N_iN_j}}\left<
\rho_i^*(\mathbf{q},0)\rho_j(\mathbf{q},0)\right>
\end{equation}
where $\tilde{h}_{ij}(q)$ denotes the Fourier transform of the total correlation function and 
$$\rho_{j}(\mathbf{q},
t)=\sum_{l=1}^{N_j}\exp\left[i\mathbf{q}\cdot\mathbf{r}_l^{(j)}(t)\right],$$
with
$\mathbf{r}_l^{(j)}$ being the coordinates of the
$l$-th particle of species $j$ ($j=1,2$) and the
asterisk denoting the complex conjugate.
The partial static structure factors calculated within
RY approximation for binary star polymer mixtures have
been found in
very good agreement with those obtained by simulations
employing the effective potential of Eq.\
(\ref{eq:pot})
\cite{mayer07}.

Our theoretical study of the glass transition is based
on the ideal Mode
Coupling Theory (MCT), a theory that describes the
time evolution of
the density auto-correlation functions, starting only
from the knowledge of the static structure factors,
via a set of coupled
integro-differential equation \cite{goetze91}.
The normalized time-dependent collective density
autocorrelation functions are defined as,
\begin{equation}
\phi_{ij}(q,t)=\left<\rho_i^*(\mathbf{q},0)\rho_j(\mathbf{q},t)\right>/\left<\rho_{i}^*(\mathbf{q},
0)
\rho_{j}(\mathbf{q}, 0)\right>.
\end{equation}
It is also useful to focus on the self part of the
density
correlation functions, which describes the dynamics of
a tagged particle,
\begin{equation}
\phi^s_{j}(q,t)=\left<\sum_{l=1}^{N_j}\exp\left\{i{\bf
q}\cdot\left[{\bf r}_l^{(j)}(t)-{\bf
r}_l^{(j)}(0)\right]\right\} \right>.
\end{equation}

The  long-time limit values of $\phi_{ij}(q,t)$, i.e.\
 the partial non-ergodicity factors, respectively for
the collective and self
correlation functions, are defined as,
\begin{eqnarray}
f_{ij}(q)=\lim_{t\rightarrow\infty}\phi_{ij}(q,t)\\
f^s_{j}(q)=\lim_{t\rightarrow\infty}\phi^s_{j}(q,t).
\end{eqnarray}
A glass transition is identified within MCT as an
ergodic to non-ergodic transition,
when the non-ergodicity factor
discontinuously jumps from zero, typical of a fluid,
to a finite value typical of a glass \cite{goetze91}.

In binary mixtures with large size asymmetries,
there is a vast separation of time-scales between
large (species 1) and  small (species 2) stars due to
the enormous
difference in mobilities.
Therefore we can use
one-component MCT for the large stars only, and solve
the equation for the non-ergodicity parameter
$f_{11}(q)$ \cite{Zacca_PRL},
\begin{equation}
\label{eq:mctone}
\frac{f_{11}(q)}{1-f_{11}(q)}=\frac{1}{2}\int\frac{d^3k}{(2\pi)^3}V(\mathbf{q},\mathbf{k})f_{11}(k)f_{11}(|\mathbf{q}-\mathbf{k}|),
\end{equation}
with
\begin{eqnarray}
\nonumber
V(\mathbf{q},\mathbf{k})& =
&\frac{\rho_1}{q^4}\left[\mathbf{q}\cdot(\mathbf{q}-\mathbf{k})
\tilde{c}_{11}(|\mathbf{q}-\mathbf{k}|)+\mathbf{q}\cdot\mathbf{k}\tilde{c}_{11}(k)\right]^2
\\
&\times&
S_{11}(q)S_{11}(k)S_{11}(|\mathbf{q}-\mathbf{k}|),
\label{kernel:eq}
\end{eqnarray}
The solution with the largest $f_{11}(q)$ is the
long-time limit
\cite{goetze91} of the density correlators.  We note
here that, in the
one-component effective treatment, we use $S_{11}(q)$
calculated from
solving the binary OZ equation within RY closure,
arising from the
full binary interactions of Eq.\ (\ref{eq:pot}). In
this respect, we
are treating the small stars as an effective medium,
and their
influence on the interactions between the large stars
is taken into
account explicitly and not through an effective
one-component picture
\cite{mayer04}.

MCT assumes a vast separation of time scales
between the slow
and fast degrees of freedom. For small size
asymmetries,
the separation of timescales between the different
star species
decreases and one-component MCT cannot be used
\cite{Zacca_PRL}.
In the general case of a mixture, when relaxation
time-scales of the two species are comparable, the
long-time MCT equations (Eq. \ref{eq:mctone}) are
easily generalized \cite{gotze03,barrat90} in terms of
matrices as,
\begin{equation}
\mathbf{\bar{f}}(q)=\mathbf{\bar{S}}(q)-\left\{\mathbf{\bar{S}}(q)^{-1}+
\mathcal{F}([\mathbf{\bar{f}}],q)\right\}^{-1},
\label{eq:mct2}
\end{equation}
where $\mathbf{\bar{S}}(q)=\sqrt{x_ix_j}S_{ij}(q)$,
with $x_i =
N_i/(N_1 + N_2)$ being the num\-ber
con\-cen\-tra\-tion of spe\-cies
$i$.

The functional $\mathcal{F}([\mathbf{\bar{f}}],q)$ is
defined as
\begin{eqnarray}
\nonumber \mathcal{F}_{ij}([\mathbf{\bar{f}}],q) &=&
\frac{1}{2q^2}\frac{\rho}{x_ix_j} \sum_{mnm^\prime
n^\prime}\int\frac{d^3k}{(2\pi)^3}
V_{imm^\prime}(\mathbf{q},
\mathbf{k}) \nonumber \\ &\times& f_{mn}(k)f_{m^\prime
n^\prime}(p)V_{jnn^\prime}(\mathbf{q}, \mathbf{k}),
\label{tensor:eq}
\end{eqnarray}
with $\rho = \rho_1 + \rho_2$ and $p=|\mathbf{q}-\mathbf{k}|$.
The vertices $V_{imm^\prime}(\mathbf{q}, \mathbf{k})$
depend on the equilibrium structure of the system and
are given by
\begin{eqnarray}
\nonumber
V_{imm^\prime}(\mathbf{q},
\mathbf{k})&=&\frac{\mathbf{q}\cdot\mathbf{k}}{q}\tilde{c}_{im}(k)
\delta_{im^\prime}\\
\nonumber
&+&\frac{\mathbf{q}\cdot(\mathbf{q}-\mathbf{k})}{q}
\tilde{c}_{im^\prime}(|\mathbf{q}-\mathbf{k}|)\delta_{im}.
\end{eqnarray}
In the case of binary or multicomponent MCT, the
theory predicts,
through the couplings between species in Eq.\
(\ref{eq:mct2}) a
simultaneous jump from zero to a finite value for all
partial
collective non-ergodicity parameters $f_{ij}(q)$.
However, this does
not hold for the self part $f^s_{i}(q)$, for which the
equations are
\cite{thakur91b},
\begin{equation}
\begin{aligned}
K_i(q)&=\frac{1}{\rho_iq^2}\sum_{j,k}\int\frac{d^3q^\prime}{(2\pi)^3}f_i^s(|\mathbf{q}-\mathbf{q^\prime}|)\tilde{c}_{ij}(q^\prime)\tilde{c}_{ik}(q^\prime)\\
&\times\sqrt{S_{jj}(q^\prime)S_{kk}(q^\prime)}f_{jk}(q^\prime),
\end{aligned}
\end{equation}
where
\begin{equation}
f_i^s(q)=\frac{1}{1+q^2/K_i(q)},
\end{equation}
so that it is possible to distinguish also the case of
mobile
particles in a frozen environment, for example for the
small stars in
our case, when $f_{22}(q)\ne0$ but $f^s_{2}(q)=0$
\cite{thakur91a,
thakur91b}.

In general multiple glassy solutions of Eq.\
(\ref{tensor:eq}) can
exist, due to higher order bifurcation singularities.
This case has
been extensively studied for effective one-component
systems of
attractive colloids, where higher order MCT
singularities have been
predicted theoretically \cite{Daw00a} and verified by
numerical
simulations \cite{Scio03,Zac03a} and experiments
\cite{pham,bartsch,Gra04a}.  For multicomponent
systems, calculations
of MCT higher order singularities have not been
reported so
far. However, hints of the presence of such higher
order singularities
and glass-glass transitions, in particular of one of
the involved
species, have emerged in simulations of asymmetric
polymer blends and
soft sphere mixtures
\cite{Moreno06JCP,Moreno06PRE,Moreno06condmat}.
 In the
case of distinct glassy states, the non-ergodic
properties  of the various arrested states will be
different from one another, a feature that will
be reflected on the corresponding non-ergodicity 
factors $f_{ij}(q,t)$.  Also, close
to such singularities, competition between the
different glassy states
modifies the standard MCT dynamical behavior close to
the liquid-glass
transition (labeled usually of type $B$ or $A_2$). For
such an $A_2$
transition, a two-step relaxation is well described by
MCT, through an
asymptotic study of the correlators near the ideal
glass solutions.

For a generic density correlator $\phi$ in the liquid the
critical
non-ergodicity parameters $f^c(q)$ can be extracted.
The departure from the plateau,
i.e. the start of the $\alpha$-process, is described
by a
power law, called the von Schweidler law, regulated by
the
characteristic exponent $b$, which is independent of
the particular $q$-vector considered,
\begin{equation}
\phi(q,t)- f^c(q) \sim -h^{(1)}(q)
(t/\tau_0)^{b}+h^{(2)}(q)(t/\tau_0)^{2b}
\label{eq:vs-law}
\end{equation}
with $\tau_0$ being the characteristic time of the
relaxation.
The quantities $h^{(1)}(q)$ and $h^{(2)}(q)$ are referred to
respectively as critical
amplitude and correction amplitude \cite{goetze91}.
The $\alpha-$relaxation process can  also be
well
described by a stretched exponential, i.e.
\begin{equation}
\phi(q,t)=A(q)\exp{[-(t/\tau_q)^{\beta_q}]}
\label{stretched}
\end{equation}
where the amplitude $A(q)$ determines the plateau value, and the
stretching exponent $\beta_q\leq1$.  For large $q$ values, it has been
shown that $\beta_q\to b$ \cite{fuchs}. The von Schweidler exponent
$b$ is related to another important MCT parameter, named the exponent
parameter $\lambda$\cite{goetze91}. This includes all information on
the dynamics of the systems on approaching the MCT transition.  It
takes values in the interval $1/2 < \lambda \leq 1$.

Close to higher order singularities, the dynamical behavior is
predicted to obey a logarithmic behavior and $\lambda$ should approach
$1$.  In particular, it holds,
\begin{equation}
\label{eq:mct-log}
\phi(q,t)\sim f^c(q)-h(q) \left[B^{(1)} \ln(t/\tau) +
B^{(2)}_q
\ln^2(t/\tau)\right].
\end{equation}
We define $q^*$ as the wave-number at which the correlators decay
purely logarithmically, that is $B_{q^*}^{(2)}=0$.  Here $\tau$ stands
for a time-scale which diverges if the state approaches the
mathematical singularity. The formula is obtained by asymptotic
solution of the MCT equations \cite{gotzesperl}. The first term
$f^c(q)$ is the sum of the non-ergodicity parameter at the singularity
plus a correction which depends on the distance from the singularity.

We will use Eq.~(\ref{eq:vs-law})--(\ref{eq:mct-log})
to describe the  the correlators $\phi_{11}(q,t)$ and $\phi_{22}(q,t)$
independently.
In the following, for those states where
the density correlators do not follow
power-laws or stretched exponentials, we adopt
the expression in
Eq.\ (\ref{eq:mct-log}) to extract the non-ergodicity
parameters from
the simulations and find that they are indeed very well described by that function.

\section{MCT Results}

Following the experimental setup discussed in \cite{Zacca2005},
we investigate the effect of the addition of a second component of varying size and concentration on the glass formed by
the one-component large stars with $f_1=263$.
As we have shown before \cite{Zacca2005}, the qualitative trends do not depend
on details of the mixture properties, therefore we consider only $f_2=64$ in this work.
We commence from a glass of large stars only, with a fixed density $\rho_1\sigma_1^3=0.345$,
corresponding to a glassy state within MCT
\cite{foffi03}.   At this state point, we add small stars and study
 the stability of the large star glass
\footnote{MCT long-time limit equations are solved iteratively discretizing the
wave vector integrals on a grid of 1000 points for the binary equations (Eq.\ref{eq:mct2})
and 1600 for the one-component ones (Eq.\ref{eq:mctone}), with a mesh in wave vectors
respectively of $0.1/\sigma_1$ and $0.0625/\sigma_1$.}.

\begin{figure}
\centering
\includegraphics[width=8cm, clip =true]{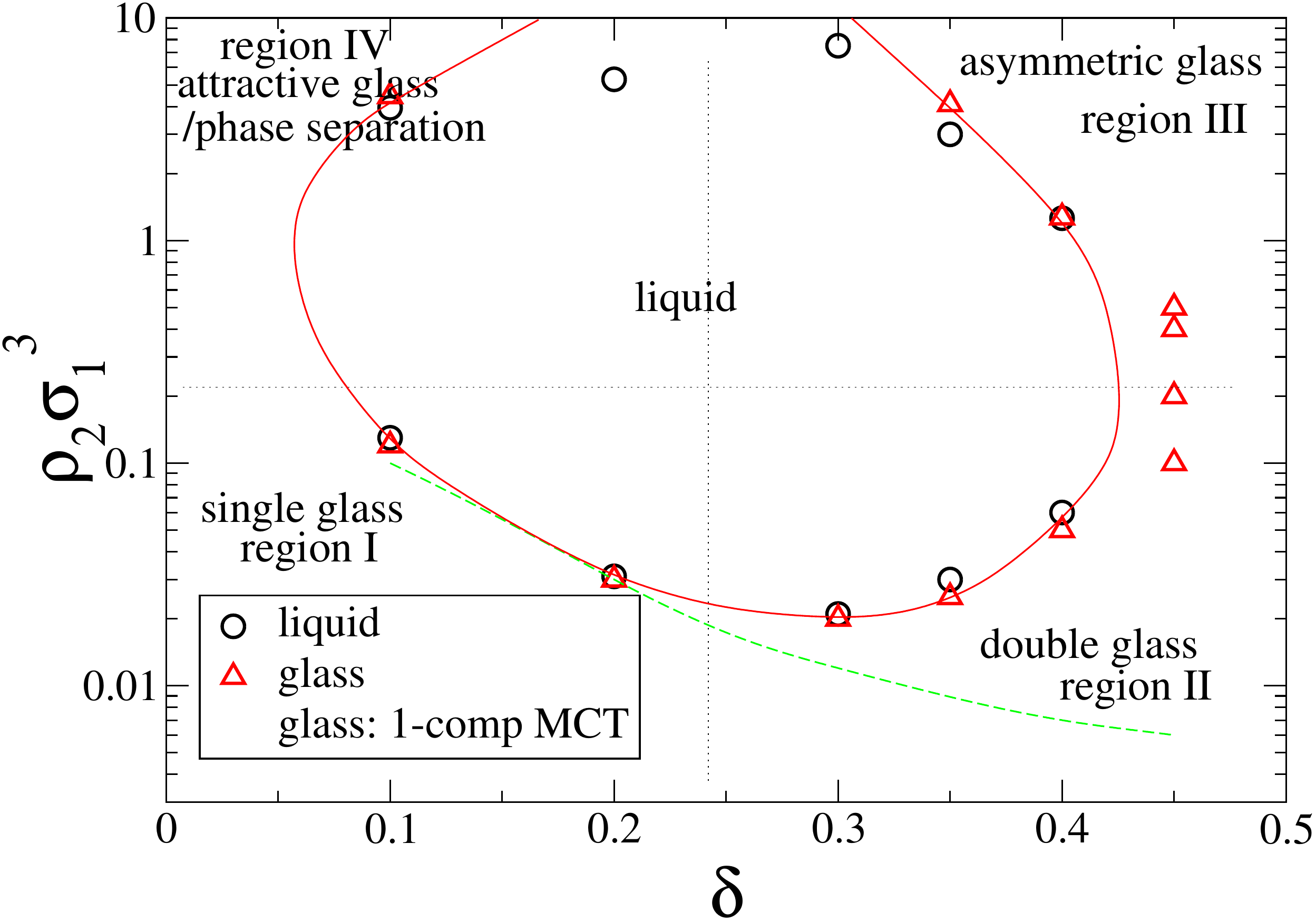}
\caption{Ideal MCT kinetic phase diagram in the size ratio-additive
  density plane.  Symbols are the bracketing state points for the
  ideal-glass transition, determined by using the two-component MCT.
  The full line through the points is a guide to the eye. The dashed
  line is the ideal-glass line calculated on the basis of the
  one-component MCT.  Four different regions with glassy states can be
  identified: we name them respectively single glass (region I),
  double glass (region II), asymmetric glass (region III) and
  attractive glass/phase separated glass (region IV).  These
  calculations refer to the case $\rho_1\sigma_1^3=0.345$.  The dotted
  lines roughly separate the different regions of the phase diagram.}
\label{fig:mctphase}
\end{figure}

From the MCT analysis, we determine a glass transition diagram in the
$(\delta,\rho_2)$ plane (see Figure \ref{fig:mctphase}). 
Results obtained both in the full binary treatment of the
mixture and in the effective one-component representation are
shown.  
With respect to the $(\rho,T)$ phase diagrams that are usually shown
in colloid-polymer (CP) mixtures \cite{pham}, we focus here on a single isochore
and consider the addition of small stars of  different sizes and concentrations.

Examining Fig.\ \ref{fig:mctphase}, a liquid region is found, being
surrounded by many different glasses. We detect a reentrant behavior
on increasing the density of additives both for small and large sizes
of additives as well as a reentrance in size ratio.  For convenience,
we can roughly divide the phase diagram of star-star mixtures in four
different regions to be discussed in the following:
\begin{enumerate}
\item small size and small concentration of the additives (region I):
  single glass;
\item large size and small concentration of the additives (region II):  double glass; 
\item large size and large concentration of the additives (region III):  asymmetric glass; 
\item small size and large concentration of the additives (region IV):  attractive glass/phase separation.
\end{enumerate}

Preliminary information on the single, double and asymmetric glasses have been
reported in Ref.~\cite{Zacca2005,unpublished}.  Here we summarize the most important aspects for completeness.  We start by examining region I. 
The small stars always remain liquid in this regime, even in
those states in which the large ones are arrested (single glass)
 \cite{Zacca2005}.  The addition of small stars with
size ratio up to $\delta \approx 0.25$ causes the large glass star to melt due
to a depletion-induced softening of the repulsions with 
respect to the one-component case \cite{mayer04}, weakening the stability of the cages \cite{Zacca2005}.

On the other hand, moving into region II by increasing the small stars diameter, 
the addition of a second
component of  stars with $0.25 \lesssim \delta \lesssim 0.45$ leads to
a glassy state formed by both components (double glass) \cite{Zacca2005}.
Here, an increase of $\rho_2$ also
melts the glass but through an entirely
different mechanism, found in MCT and also confirmed by
experiments \cite{Zacca2005}. Indeed, the glass becomes stiffer through
the addition of a second glassy component, as shown by the growth of
the elastic modulus in theory and experiments reported in
\cite{Zacca2005,mayer07} up to the melting density. At this point, it becomes
entropically more convenient  to free available volume for both stars
simultaneously, which amounts to a finite energetic increase, sufficient to cause liquification. 
Therefore, this melting mechanism is only possible
through the softness of interactions involved, and it does not have an
equivalent in terms of hard-sphere mixtures.  We note, that the presence of two
different glasses for asymmetric mixtures, a single to a double
glass, depending on the size ratio,
was already found for
hard spheres \cite{thakur91a,thakur91b,imhof95}, but the new feature of
the presently investigated soft mixtures is that both glasses can be
simultaneously melted by increasing the amount of the small additives
in the mixture.

Clearly, a theoretical framework capable to correctly describe the
single glass should be based on the slowing down of the large stars
only, i.e.\ one-component MCT, where the smaller stars are assumed to
form a fluid medium that does not participate in the glass formation,
is the conceptually right framework to use. On the other hand, in the
double glass both species become comparably slow, thus compelling the
use of binary MCT.  It turns out that in the case of star polymers
mixtures, differently from that of colloid-polymer mixtures
\cite{Zacca_PRL}, at low $\delta$, i.e., in the single glass regime,
both treatments give qualitatively the same behavior, as illustrated
also in Figure \ref{fig:mctphase}, while at large $\delta$, in the
double glass regime, the discrepancy becomes large, signaling the
tendency of the small component to also arrest. Only binary MCT can be
used in this region, giving rise to the U-shape also observed in the
experiments for low $\rho_2$ \cite{Zacca2005}.  Hence, in the
following we just refer to two-component MCT in all regions, unless
explicitly mentioned.

In the present work we complement previous results \cite{Zacca2005,
  mayer07} with the investigation of Regions III and IV.  In region
III, for large concentration of additives with $\delta\sim 0.4$, we
find a reentrant vitrification with respect to $\rho_2$, with the
appearance of a new glassy state with smaller localization length for
large $\delta$ and large $\rho_2$.  Due to the softness of the
interactions local rearrangements take place which leads to a change
of coordination of the large stars as well as to an asymmetric
deformation of the cages \cite{unpublished}.  The increase of $\rho_2$
causes a second jamming at a new length scale that will depend on the
particular parameters of the mixture.

Finally in region IV, at low $\delta$, the depletion mechanism induces
effective attractions between the large stars \cite{mayer04}, causing
phase separation.  Phase separation is accompanied by an increase of
the structure factor at infinite wavelength, and a blind use of it in
the mode-coupling equations would predict glass formation driven by
the small wavevectors.  However, with the use of RY structure factors,
we are only able to find a glass within binary MCT for $\delta=0.1$,
while the one-component treatment always provides a liquid state up to
the maximum investigated $\rho_2$ approaching phase separation. We
could not extend the analysis to larger $\rho_2$ due to a breakdown in
the convergence of the Rogers-Young closure, as well as the simplest
HNC closure in this region, a possible sign of a close thermodynamic
instability.

\begin{figure}
\centering
\includegraphics[width=8.5cm,clip]{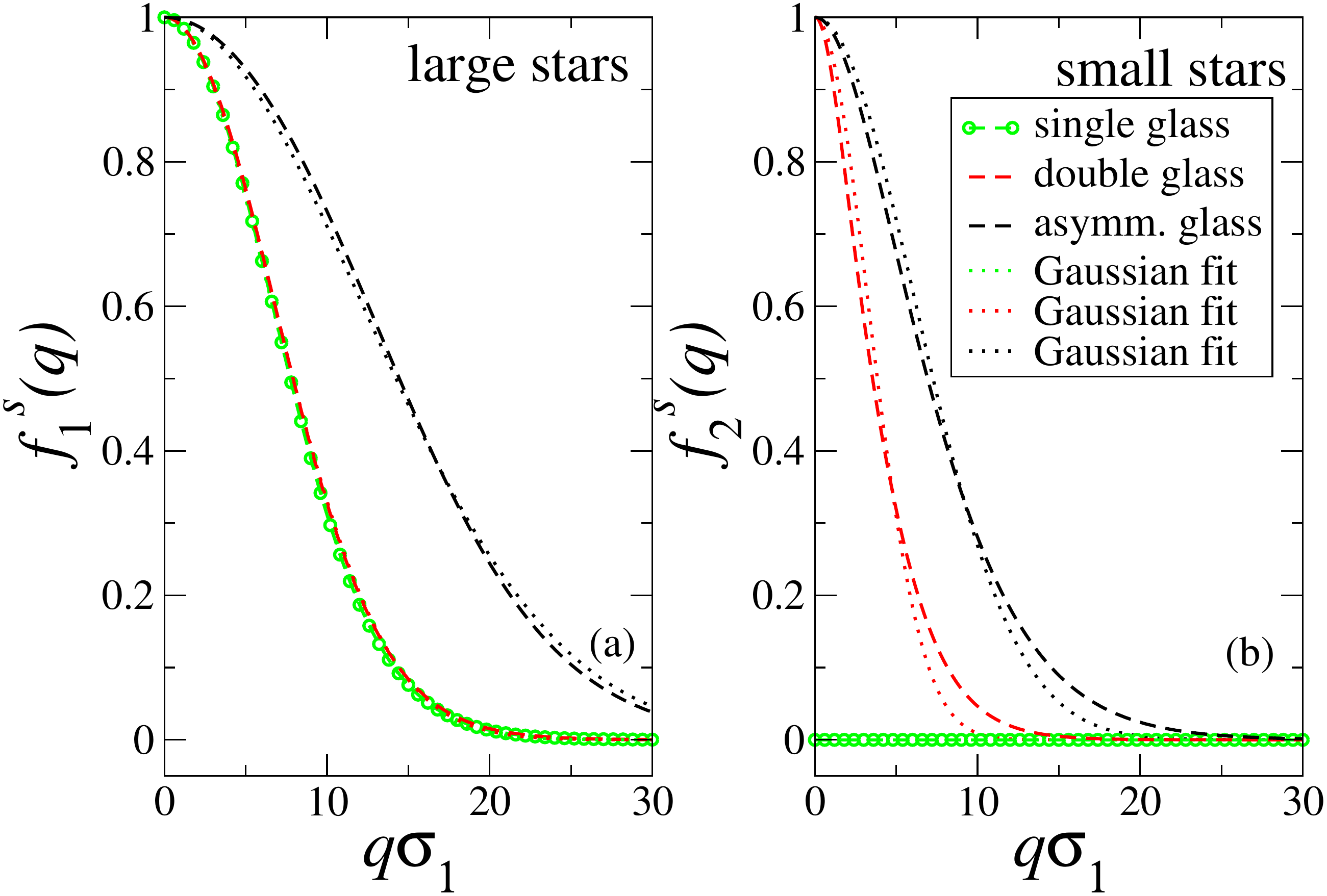}
\includegraphics[width=8.5cm,clip]{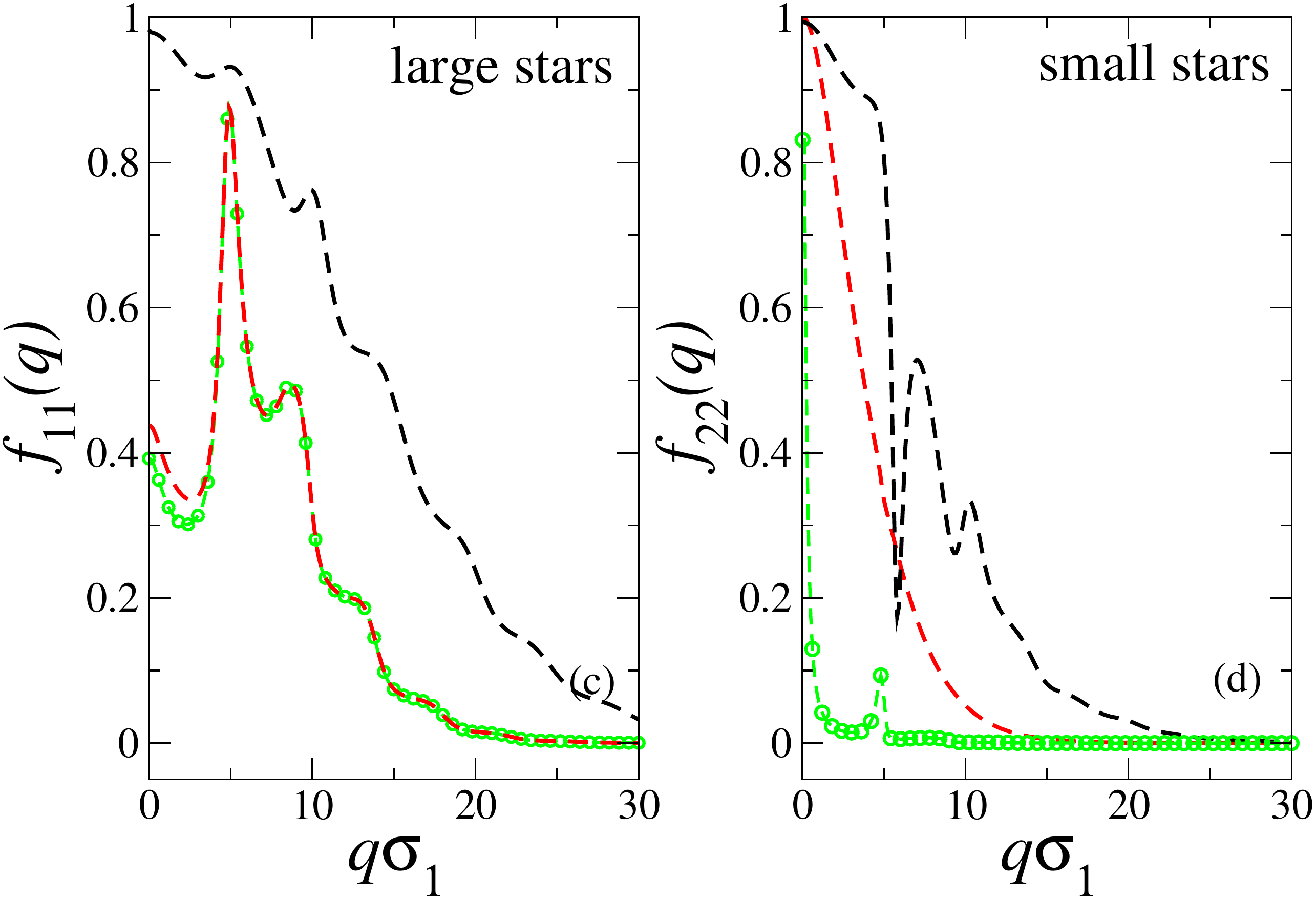}
\caption{Partial non-ergodicity parameters at the MCT transition. 
The (a) and (b) panels show respectively the large-star and small star components of the self non-ergodicity parameters 
 for the three different glasses. They also show  Gaussian fits (dotted lines), which allow us to extract
the localization length.
The $f_1^s$ of the large stars in the single and double glass are virtually identical, therefore the lines fall on top of each other, 
while those for the small stars are very different
in all three cases.
 In the (c) and (d) panels  we show the non-ergodicity parameters  for the collective dynamics.
For all curves $\rho_1\sigma_1^3=0.345$. For the single glass
 $\rho_2\sigma_1^3=0.12$ and $\delta=0.1$, 
while in the double glass  $\rho_2\sigma_1^3=0.05$ and $\delta=0.4$. The asymmetric glass corresponds to 
$\rho_2\sigma_1^3=1.27$ and $\delta=0.4$. 
}
\label{fig:mct-fq}
\end{figure}

In addition to the phase diagram, we report in Fig.~\ref{fig:mct-fq}
the calculated MCT partial non-ergodicity parameters, both for self
[Fig.~\ref{fig:mct-fq}(a-b)] and collective
[Fig.~\ref{fig:mct-fq}(c-d)] relaxation dynamics for $\delta=0.1$ and
$\delta=0.4$.  Such results correspond to the glassy points closest to
the transition, hence they are `critical' non-ergodicity parameters.
The small stars $f_{2}^s(q)$ allows us to distinguish immediately a
single from a double glass, since it is identically zero for the
single glass, as already found in binary asymmetric hard sphere
mixtures \cite{thakur91b}. Thus, although, in binary MCT the broken
ergodicity transition is postulated to happen simultaneously for both
species, the self non-ergodicity parameter allows to distinguish
between mobile and immobile species.  Also, the partial $f_{i}^s(q)$
for both types of stars have a larger width in the asymmetric glass
than in the double glass.  Fitting $f_{i}^s(q)$ to a Gaussian curve
$\exp(-q^2/6l_\mathrm{MCT}^2)$ provides an estimate for the
localization length $l_\mathrm{MCT}$ within the glass. Fits are also
reported in the figure. The Gaussian fit works very well for large
stars, but deviations are observed for small stars, in both double and
aymmetric glasses. The extracted localization length is
$l_\mathrm{MCT}\approx 0.26\sigma_1$ for large stars both in single
and double glass, while it is $l_\mathrm{MCT}\approx 0.14\sigma_1$ for
the asymmetric glass, i.e., the cage size reduces by a factor of two.
A larger variation in $l_\mathrm{MCT}$ is predicted for the small
stars from $0.51\sigma_1$ for the double glass at low $\rho_2$ to
$0.11\sigma_1$ for the asymmetric glass at high $\rho_2$, a value
which is almost identical to the one for the large stars in the same
state.  For the collective $f_{ii}(q)$, the results are very similar
to those of the self non-ergodicity parameters, only much stronger
oscillation peaks are observed in phase with the static structure
factor.  For the small stars, these appear only in the asymmetric
glass regime.  We notice that upon changes in $\delta$ within the low
$\rho_2$ glasses, we detect no change in $f_{11}(q)$ between single
and double glass. However, a significant increase is observed for
large $\delta$, upon increasing $\rho_2$, while approaching the
asymmetric glass.  This increase results in a more extended $q$-width
in $f_{11}(q)$, a finding similar to that for attractive glasses in
colloid-polymer mixtures, where further increase of the additives
leads to an increase of the non-ergodicity parameter both in amplitude
and range \cite{Zac01PRE}.  Therefore, in the asymmetric regime, a
strong localizing mechanism is at hand, with some similarities to an
effective attraction. However, looking at the partial $S_{ij}(q)$ we
do not see an increase in the compressibilities or a non-monotonic
dependence in the growth of peaks. Simply, the structure factor peaks
for large stars continue to decrease upon addition of small stars,
while these display themselves more and more enhanced peaks.  In
\cite{unpublished}, we have shown that the change of structure is
accompanied by a dramatic decrease in the coordination number of the
large stars and an asymmetric deformation of the large star cages.

Finally, an analysis of the MCT exponents reveals that, for both
single and double glass, $\lambda \simeq 0.71$ and $b\simeq 0.625$,
the same result obtained for one-component star polymer solutions.
However, we note that $\lambda$ grows considerably close to the
asymmetric glass transition. At the transition point corresponding to
$\delta=0.4$, $\lambda \sim 0.8$, corresponding to a von Schweidler
exponent $b\sim 0.47$.

\section{Simulations: Numerical Details}

We performed standard Newtonian dynamics simulations for stars
interacting via the effective potential in
Eq.\ (\ref{eq:pot}). Although the use of Brownian dynamics simulations
would be more realistic for star-star mixtures to mimic the solvent
effects, this is less efficient and it would not allow us to probe the
full slow dynamics regime efficiently. Moreover, several confirmations
of independence of long-time properties from microscopic dynamics have
been provided so far \cite{gleim,foffi03} for systems close to their
glass line.  We investigate several compositions, in order to have a
complete picture of the phase diagram and of the dynamical slowing
down.  Due to the various compositions to be investigated, the number
of particles varies throughout the diagram. In general, we fix
$N_1=1000$ and vary $N_2$ accordingly. However, for $\rho_2\sigma_1^3
\geq 10$, we use $N_1=250$.  Also, for $\rho_2\sigma_1^3 \leq 0.1 $,
i.e.~ in single and double glass regimes, we fix the minimum number of
small particles to $N_2=500$ and vary $N_1$ accordingly.  We
investigate $f_1=263$ and $f_2=64$, the same values used in the MCT
analysis.  To improve statistics, we average over 5 independent runs
for each studied state point.

In order to describe the dynamics of the system properly, we have to
choose the short-time mobilities of the small particles with respect
to the large ones.  Following the ideas in \cite{Zacca_PRL}, we use
the scaling for the mass $m$ typical of the star polymers, namely $m
\sim f^{2/3}\sigma^{5/3}$ \cite{likos01, Grest:review:96}. This
results in a ratio between the small and large particles mass equal to
$m_2/m_1=(f_2/f_1)^{2/3} \delta ^{5/3}$. Note that a physical choice
of the small particle mass is crucial, since as already noticed in
earlier simulations of explicit colloid-polymer mixtures
\cite{Zacca_PRL}, the addition of small particles of mass identical to
the one of the large particles, has the result of further slowing down
the large particle dynamics.  This would imply no melting at any given
$\rho_2$ and $\delta$ in contrast with MCT predictions and
experimental results.  Units of length, mass and energy are
$\sigma_1$, $m_1$ and $k_\mathrm{B}T$, respectively. Time is measured
in $\tau_\mathrm{MD}=\sqrt{m_1\sigma_1^2/(k_\mathrm{B}T)}$ and the
integration time step is varied according to the mass ratio, as
$\delta\tau=5\cdot 10^{-3}\sqrt{m_2/m_1}\tau_\mathrm{MD}$.

We employ numerical simulations to (i) evaluate the iso-diffusivity
lines in the $\rho_2-\delta$ plane and (ii) study specific features of
the dynamics for selected state points approaching the four glass
regions.

Simulations for the isodiffusivity lines are performed for
$\rho_1\sigma_1^3=0.345$.  Note however that at this density the
slowing down of the dynamics observed in the simulations is not very
significant. To study the dynamics on approaching the glass
transition(s) we have been forced to increase the density of the large
particles to generate a slower dynamics. Indeed, similarly to other
studied cases like that of hard spheres, a systematic shift of the MCT
transition with respect to the slowing down taking place in
experiments or simulations, is present. This shift usually applies in
the direction of MCT anticipating the onset of glassy dynamics and our
findings agree with this general picture.  

Moreover, we have considered a small poly-dispersity in the large
particles diameter.  This is particularly important at low $\rho_2$,
when the amount of small stars is not sufficient by itself to prevent
crystallization of the large stars. We selected the value
$\rho_1\sigma_1^3=0.41$ to investigate the dynamics of single and
double glass respectively, while we retained the value
$\rho_1\sigma_1^3=0.345$ for the asymmetric glass. 
The larger density
and the absence of crystallization allows us to observe a slowing down
of the dynamics for the large particles (even in the absence of
additives), and consequently also the single and double glass.  We
employ a polydispersity in the large-stars diameter, chosen to follow
a Gaussian distribution with a width of 10\% \cite{VlaPRE05}.

\begin{figure}
\centering
\includegraphics[width=8.5cm,clip]{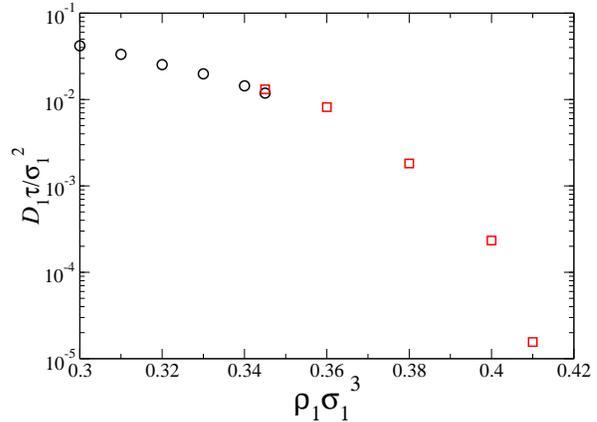}
\caption{Diffusivity of the one-component system as a function of
star density. Circles correspond to the monodisperse system with $f=263$
and squares  to simulations with added polydispersity to
prevent the system from forming a crystal. The exploration
of three orders of magnitude in $D_1$ is thus possible.}
\label{fig:onecomp}
\end{figure}

We have checked that, compared to the one-component case, the dynamics
are not affected.  For the comparison, we have to account for the
polydispersity in the right way when calculating the density, i.e.\ by
matching the value $\rho_1\left<\sigma_1^3\right>$ with the
corresponding one-component density.  In this way, we can decrease the
diffusivity by three orders of magnitude by increasing $\rho_1$, as
illustrated in Figure \ref{fig:onecomp}, comparing results for the
monodisperse and polydisperse one-component reference system.

Finally, for the simulations of the attractive glass, we have focused on
monodisperse large stars, due to the absence of systematic
crystallization in this case.

\section{Simulation: Isodiffusivity Lines}

The iso-diffusivity lines are precursors of the glass
transition lines, and they always display a similarity in shape as the
MCT line, as demonstrated already for one-component star
polymers \cite{foffi03}, for other repulsive potentials \cite{Kumar05},
as well as for short-range attractive square
well potentials \cite{Zac02a}.

\begin{figure}
\centering
\includegraphics[width=8.5cm,clip]{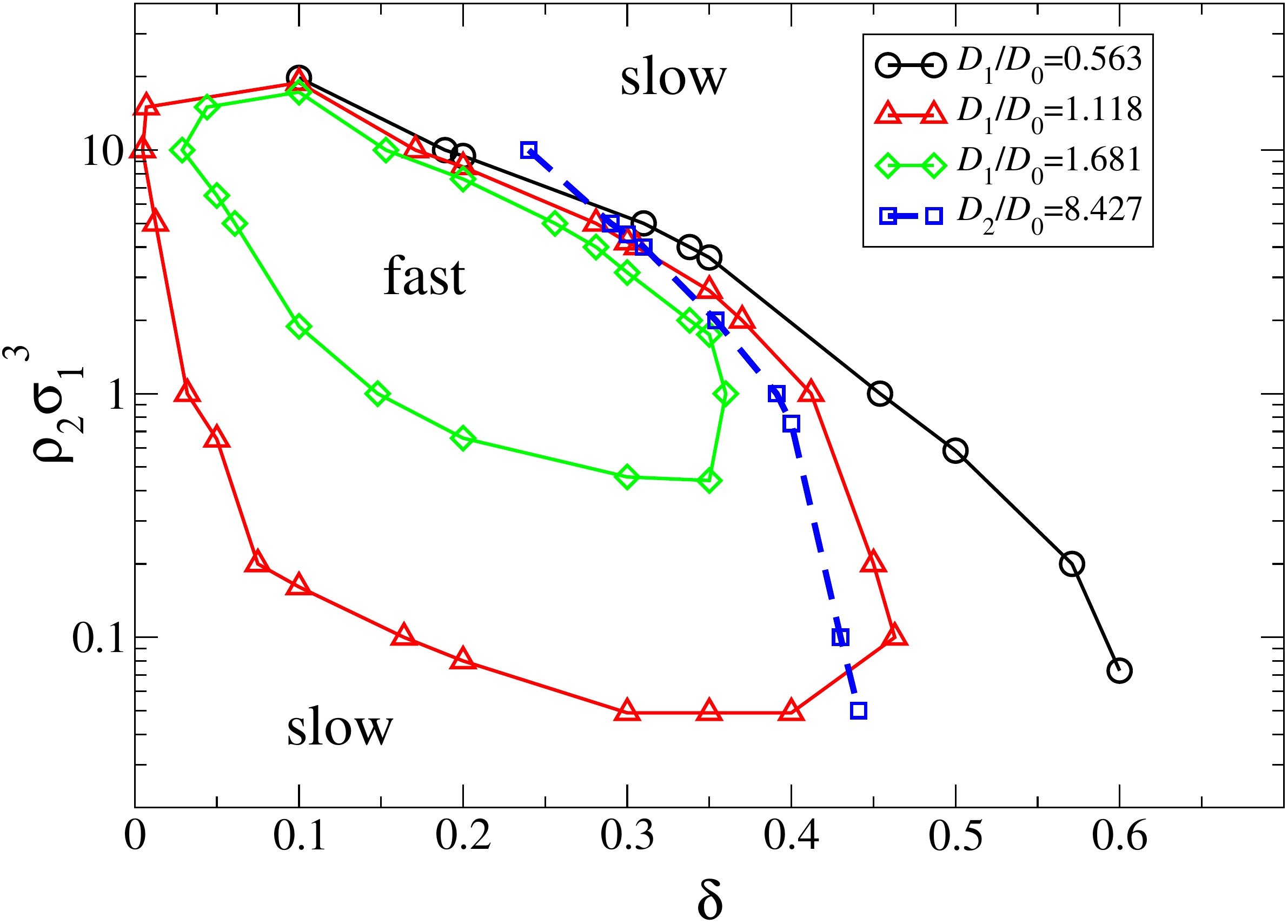}
\caption{Iso-diffusivity lines from MD simulations in the 
$(\rho_2,\delta)$-representation. The density of large stars is kept constant
at $\rho_1\sigma_1^3=0.345$ and the large stars are monodisperse. 
The line for the smallest value
of the diffusivity is not closed because the chosen value of $D_1$ is 
smaller than the diffusivity of the reference one-component system ($D_0$).
The dashed curve is an isodiffusivity line of the small stars.}
\label{fig:iso}
\end{figure}

To study iso-diffusivity lines we calculate the mean-squared
displacement (MSD) $\langle r_i^2(t)\rangle$ $(i=1,2)$ of large and
small stars separately. From the long-time behavior, we extract the
respective self-diffusion coefficient defined as, following the
Einstein relation, $D_i=\lim_{t\to\infty} \langle
r_i^2(t)\rangle/(6t)$.

In Fig.\ \ref{fig:iso}, we report the iso-diffusivity lines for three
different values of $D_1/D_0$ in the $(\delta,\rho_2)$ plane,  where $D_0$ is 
the diffusion coefficient in the absence of small particles. The fact that
we examine only a decade in decrease of $D_1/D_0$ is due to the many
constraints to study the whole phase diagram under very different
mixture compositions. 
The line
corresponding to the slowest states is drawn only on the large $\delta$
side, because the diffusivity is lower than in the one-component system
and therefore no closed loop exists for this value of $D_1$.    
Notwithstanding these details, the emergence of an
asymmetric ``O'', that resembles the shape of the MCT phase
diagram, is evident. This was confirmed also for higher $\rho_1$ \cite{unpublished}, demonstrating
that the topology does not depend on the precise value of $\rho_1$ studied.
Recently, this asymmetric ``O'' has been observed experimentally \cite{unpublished}.

Additionally, from the simulations we can also trace out correctly
the iso-diffusivity lines of the small stars (Fig.~\ref{fig:iso}). Within binary MCT, there
should be no distinction between the glass lines of the large and
small stars, based on the 
assumption that small stars are as slow as the large
stars.  It has a similar
shape as the large $\delta$/large $\rho_2$ region. Note that  the dynamics of the small stars is much faster for low $\delta$, where the mobility of the two species
is very dissimilar, and hence the small star isodiffusivity lines do not show an "O" shape.

\section{Simulation: Multiple glassy states}

We can now exploit the simulation results to directly probe the four
regions of slowing down and compare the different glasses, in order to
clarify the characteristics of the multiple glassy states appearing in
these types of mixtures, in comparison with the MCT results. 

\subsection{Region I: Single glass}

\begin{figure}
\centering
\includegraphics[width=  8.5cm,clip]{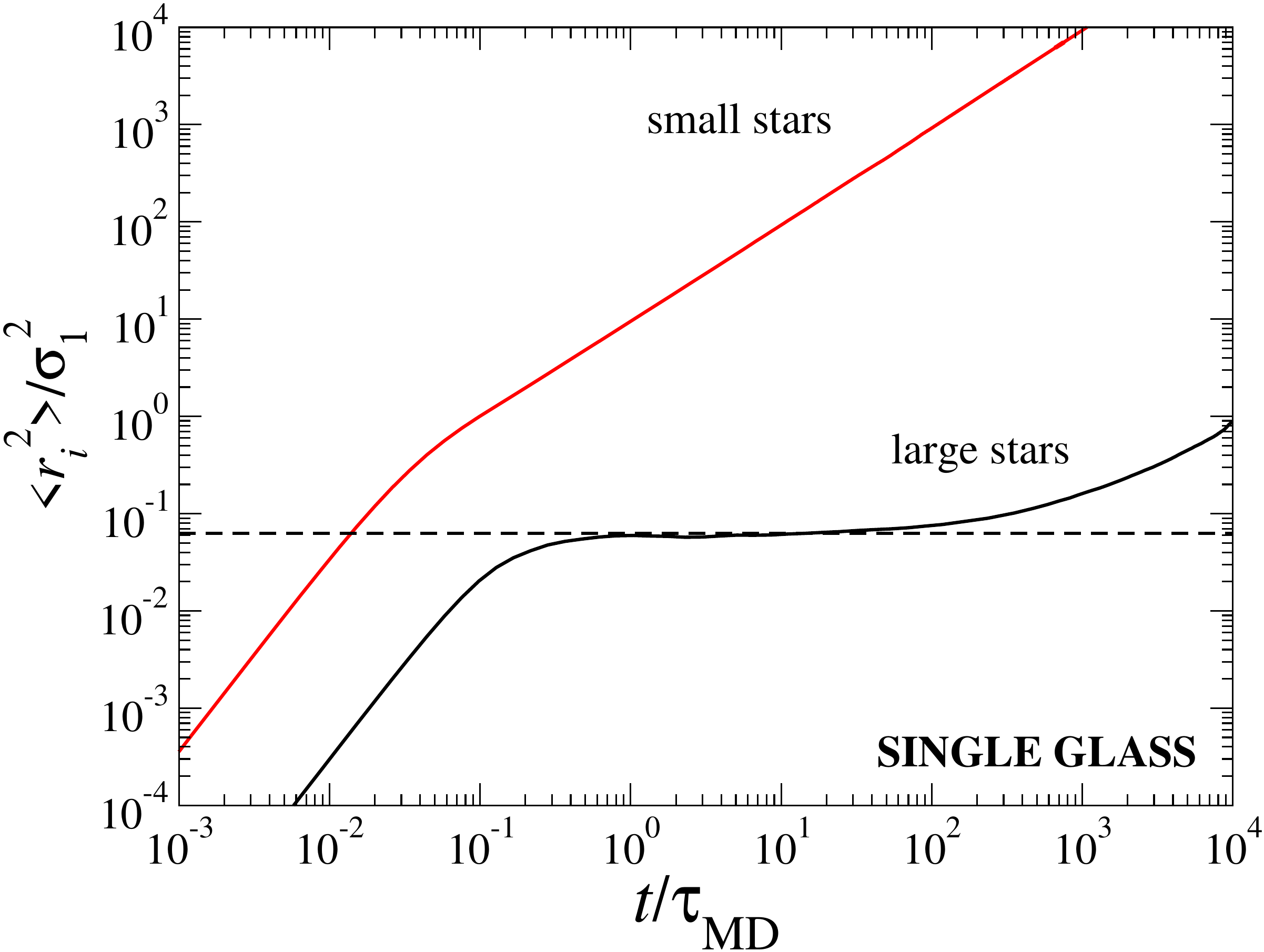}
\caption{Mean squared displacements for  large and small stars 
in the single glass. The densities are $\rho_1\sigma_1^3=0.41$
and   $\rho_2\sigma_1^3=0.1$; the size ratio is $\delta=0.1$. The horizontal line indicates
the estimate of the squared localization length for  large stars, defined as the
inflection point of the MSD.}
\label{fig:msd-SG}
\end{figure}

We consider the state point $\rho_1=0.41$,
$\rho_2=0.1$ and $\delta=0.1$ (2250 particles in the simulation box). The mass ratio is
$m_2/m_1=8.4\times 10^{-3}$, allowing an effective  one-component description
\cite{Zacca_PRL}.
Indeed, the mean squared displacement in Fig.~\ref{fig:msd-SG} shows a clear separation in
time-scales between the two species. After the initial ballistic
regime, the small stars simply become diffusive, while the large ones
display a clear plateau, lasting about three decades in time, before
eventually reaching the diffusive regime. Correspondingly, a difference in the
diffusion coefficient (and relaxation time) of about three orders of
magnitude is observed. By convention, we
define the localization length as $l_\mathrm{MD}=\sqrt{\langle
r^2(t^*)\rangle}$, where $t^*$ is the point of inflection of the
MSD in the $\log-\log$ plot. We therefore provide evidence that the large
stars are nearly arrested in a jammed state, with a localization
length that can be extracted from the plateau height
$l_\mathrm{MD} \sim 0.25\sigma_1$ (horizontal line in Figure~\ref{fig:msd-SG}).  A similar value could be extracted from a simulation of star
polymers in theta-solvent \cite{VlaPRE05}. Also our MCT
results discussed provide a similar value, suggesting rather robustly that the localization length
of a soft star glass is slightly higher than of the typical hard
sphere one due  
to the softness of the cages which allows more
flexibility in the rattling than in the case of rigid hard spheres.

\begin{figure}
\centering
\includegraphics[width=8.5cm,clip]{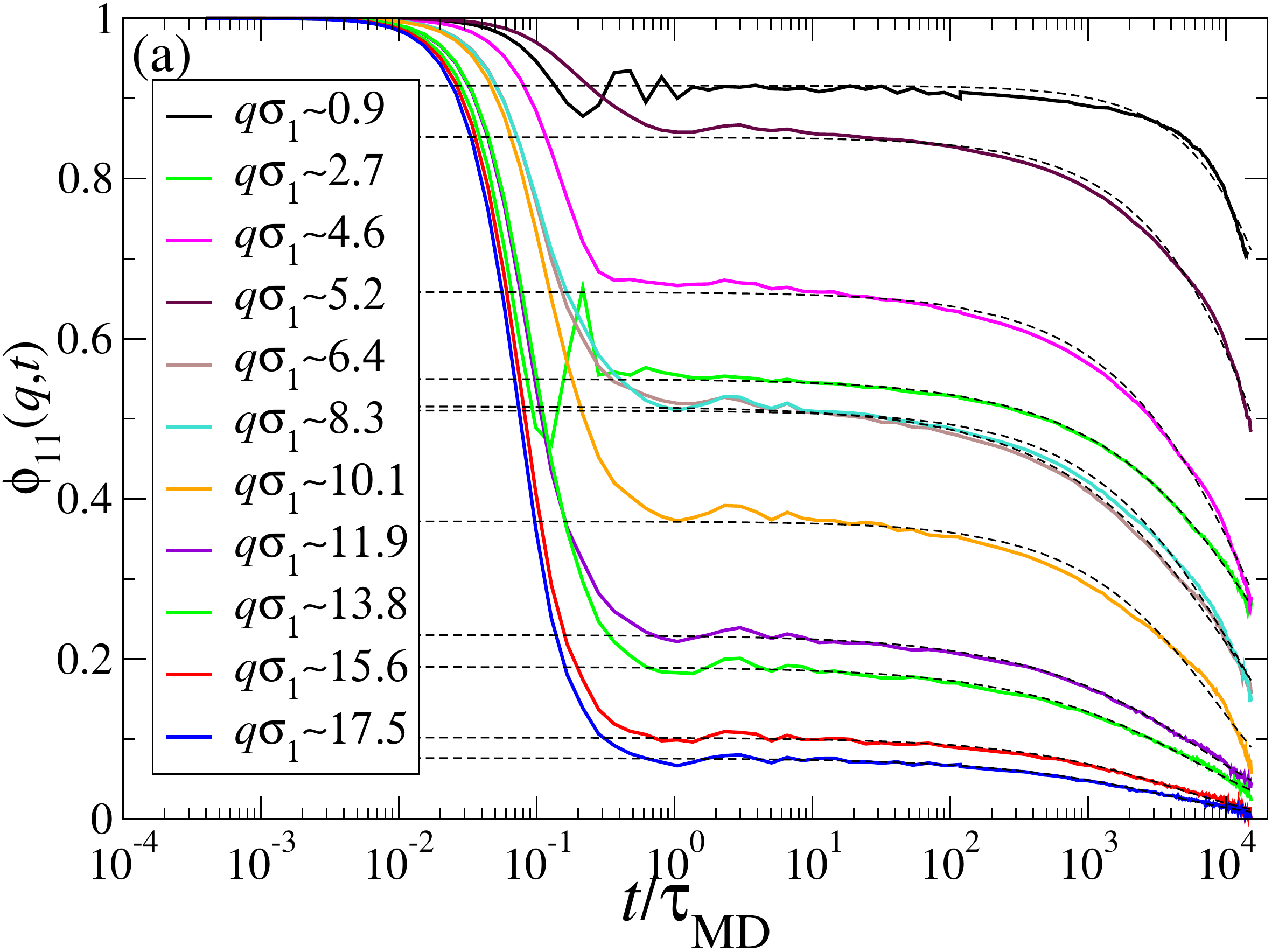}
\includegraphics[width=8.5cm,clip]{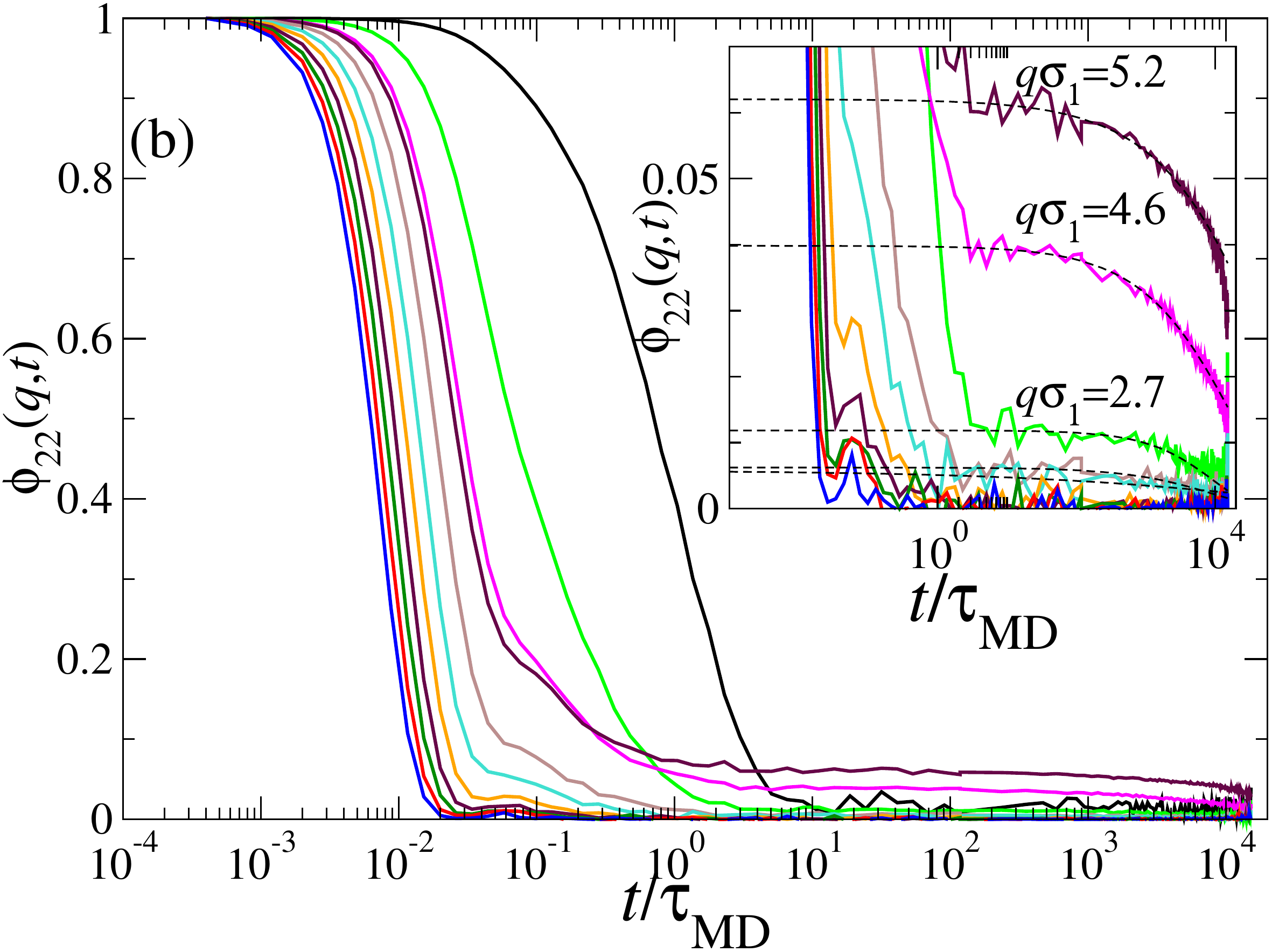}
\caption{Density correlators for  large (a) and small (b) stars
in the single class for different wave numbers $q$ (see legends).
 The parameters are the same as in Fig.~\ref{fig:msd-SG}.
The dashed lines are stretched exponential fits (Eq.~(\ref{stretched}) to the curves.
The inset (b) shows a magnification of the fits for $q$-values where small stars
also display a two-step decay.
}
\label{fig:sqt-SG}
\end{figure}

\begin{figure}
\centering
\includegraphics[width=8.5cm,clip]{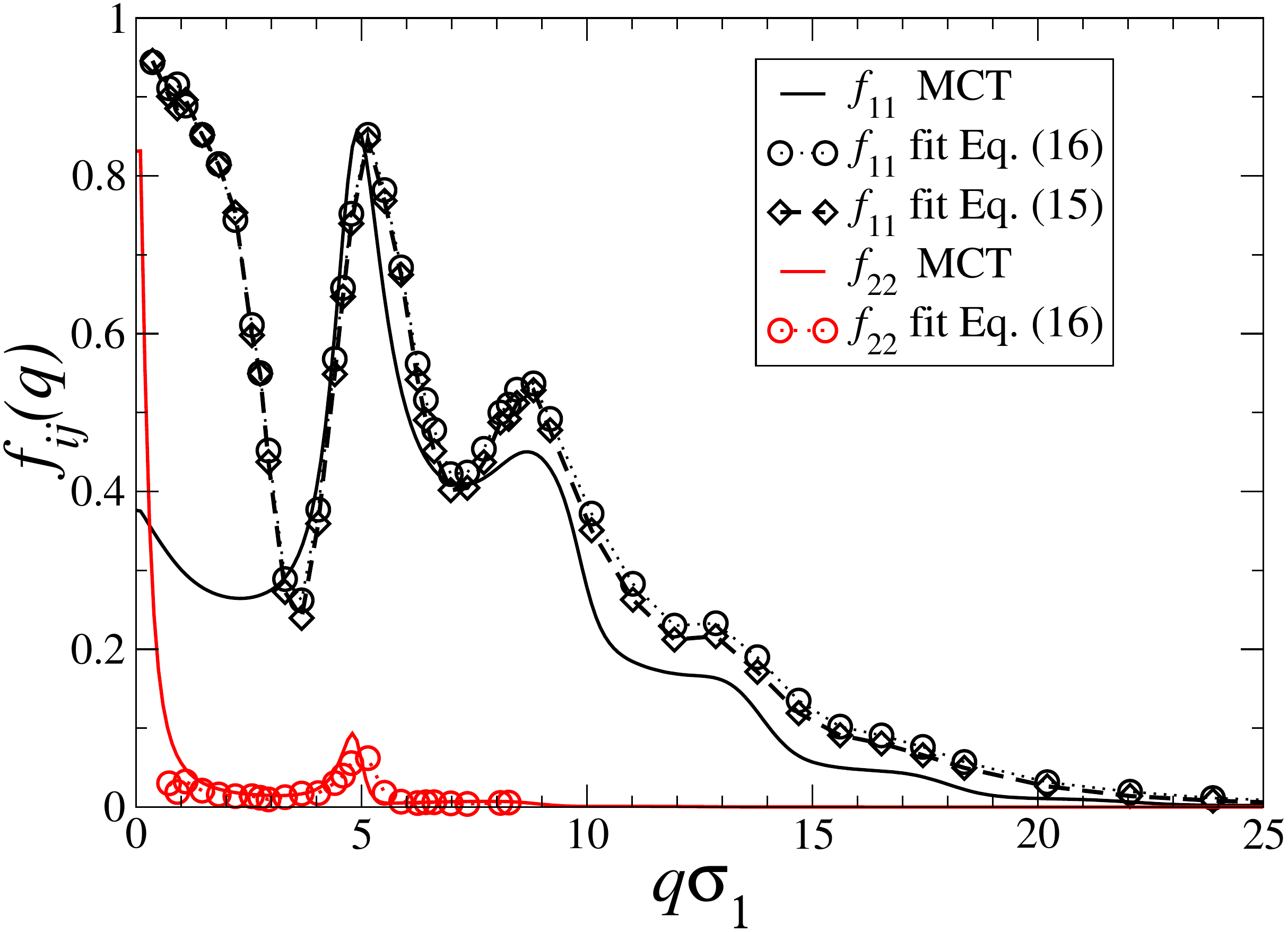}
\caption{Comparison of partial non-ergodicity parameters from MCT and
simulation in the single glass.  The values from the simulations are
obtained by stretched exponential fits (Eq.~(\ref{stretched})) for both types of stars. Also
shown are results from von Schweidler fit (Eq.~(\ref{eq:vs-law}))  for large stars with $b=0.6$.
The simulation parameters are the same as in Fig.~\ref{fig:msd-SG}.
The two component MCT results refer to the critical parameters:  $\rho_1\sigma_1^3=0.345$, $\rho_2\sigma_1^3=0.12$ and $\delta=0.1$.}
\label{fig:fqAA-SG}
\end{figure}

We now turn to examine the density correlators.  In
Fig.~\ref{fig:sqt-SG}, we show the behavior of $\phi_{ii}(q,t)$ for
large (a) and small stars (b) respectively for several wave vectors.
As for the MSD, the large stars display a marked plateau at all wave
vectors, lasting for about three decades in time.  The height of the
plateau oscillates as a function of the wave-vectors, as commonly
found in standard glasses.  We extract the non-ergodicity parameter of
the system by fitting large stars $\phi_{11}(q,t)$ with a stretched
exponential law (Eq.\ \ref{stretched}).  The results are shown in
Fig.\ \ref{fig:fqAA-SG} together with the corresponding MCT
predictions. We find that $f_{11}(q)$ oscillates in phase with the
static structure factor (not shown), but in contrast to the MCT
predictions, it shows a marked increase at low $q$, while beyond the
first peak the agreement is very good. We find that the stretching
exponent $\beta_q$ as a function of $q\sigma_1$ also oscillates in
phase with $f_{11}(q)$, and tends to approximately 0.6 for large
$q$. Indeed, using this value as $b$ exponent for the von Schweidler
law (Eq.\ (\ref{eq:vs-law})), we find that all curves are well
described in the region of departure from the plateau
(Fig-~\ref{fig:sqt-SG}(a)). This result for $b$ agrees well with MCT predictions
reported above.  The estimates for $f_{11}(q)$ from the von Schweidler
fits are in perfect agreement with that obtained from the stretched
exponential fits (Fig.\ \ref{fig:fqAA-SG}).

For the small stars correlators (Fig.\ \ref{fig:sqt-SG}(b)), we find
that the plateaus are sometimes so small that no accurate analysis can
be performed. Nonetheless at small wave vectors, the stretched
exponential fits work quite well (see inset). We find finite values
$f_{22}(q)$ just close to the first peak of $S_{11}(q)$, suggesting a
confinement of the small stars in the frozen matrix of the large
stars, which induces a small loss of ergodicity.  Thus a coupling in
time between large and small stars exists only at the peak of
$S_{11}(q)$, which might explain why two-component MCT is still
capable of correctly describing the general behavior of the mixture in
this regime.  The reason why the modes around $q \approx 5
\sigma_1^{-1}$ do not couple with modes at different $q$ values is due
to the fact that the structure factor of the small stars is
structure-less except for this $q$ value and close to the hydrodynamic
limit ($q \rightarrow 0$).

Theoretical descriptions of the dynamical arrest of
a fluid in a frozen matrix have recently been developed  \cite{krakoviack,juarezmaldonado}
which corresponds to the single glass state found here.
Fig.\ \ref{fig:fqAA-SG} shows that $f_{22}(q)$ is in good agreement with
the MCT predictions, but we do not detect any increase of
non-ergodicity at large length-scales. 
While MCT underestimates the non-ergodicity parameter of the large
stars at low $q$, it overestimates the same for the small stars
compared to the simulation results.

We can conclude from the combined study of MCT and simulations of the
single glass that it  corresponds to a glass formed by the large
stars, whose properties are those typical of 
ultra-soft star polymer glasses. The presence of the small stars  induces a weakening of the
stability of the glass, until eventually melting it. We observe this
in the behavior of the large stars' diffusion coefficient which
grows with increasing density of the additives. The small
stars are truly ergodic and fully mobile within the voids of the glassy matrix. They are never trapped in a metastable glass
state. Nevertheless, they show a  loss of ergodicity for length-scales of
the order of the static structure factor first peak for the large
stars. We remark that this behavior is seen both in binary MCT and simulations, 
while it could not have been detected by an effective one-component treatment of large stars only.

\subsection{Region II: Double glass}

We now turn to examine the arrested state for low $\rho_2$ at large
$\delta$. We consider the case $\rho_1\sigma_1^3=0.41$, 
$\delta=0.4$ and $\rho_2\sigma_1^3=0.1$, so
that the composition of the mixture is the same as for the single
glass examined above. However, the additives are larger and,
consequently, also their mass is larger.
Here the mass ratio is
$m_2/m_1=8.5\cdot 10^{-2}$, which is one order of magnitude larger than in the single glass. 
Indeed, the simulations confirm an arrest of both species.

\begin{figure}
\centering
\includegraphics[width=8.5cm,clip]{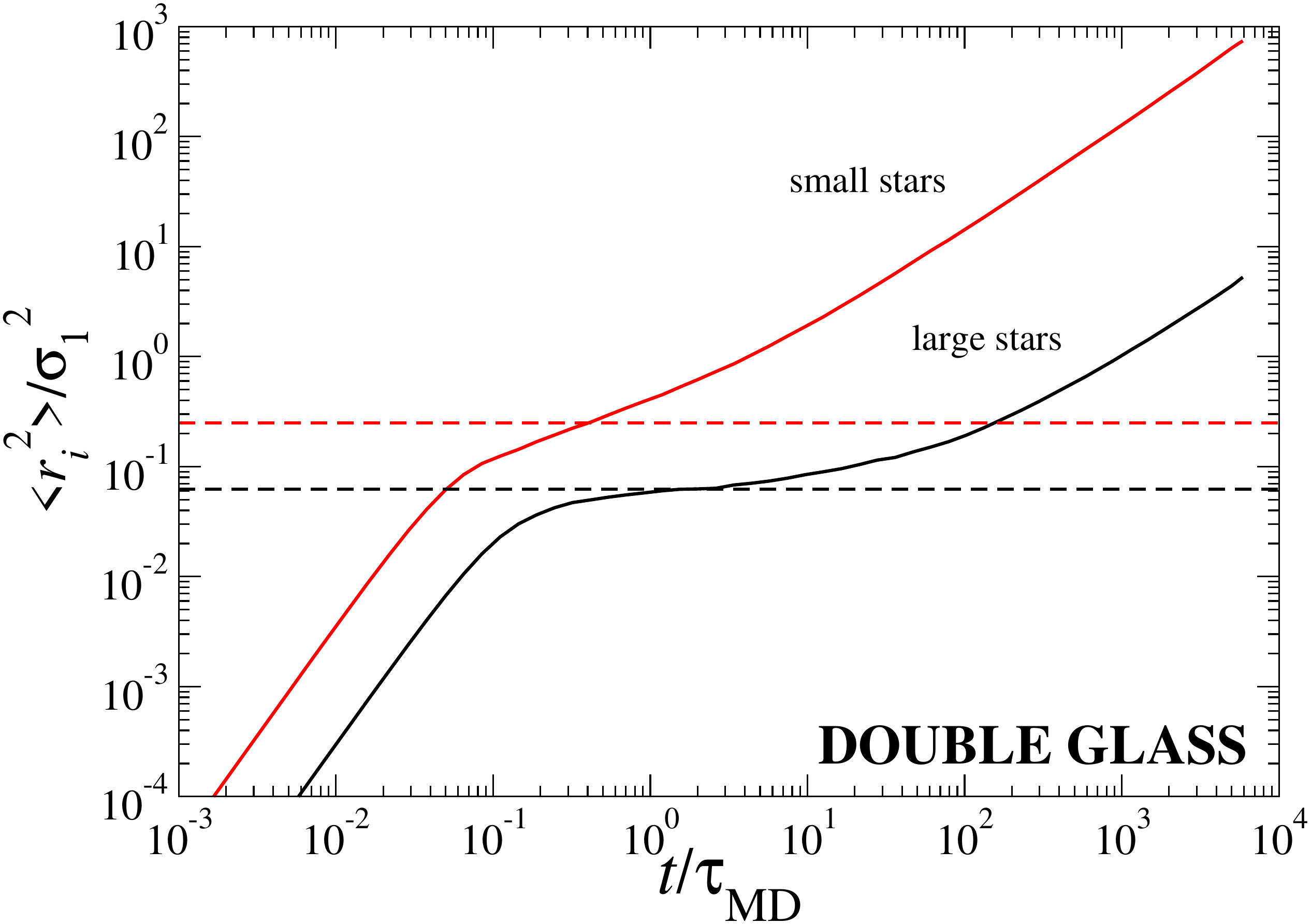}
\caption{Mean squared displacements for  large and small stars 
in the double glass. The densities are $\rho_1\sigma_1^3=0.41$
and   $\rho_2\sigma_1^3=0.1$; the size ratio is $\delta=0.4$. The horizontal lines indicate
the estimates of the squared localization length for  large and small stars, respectively, defined as the
inflection point of the MSD.}
\label{fig:msd-DG}
\end{figure}

In Fig. \ref{fig:msd-DG}, we show the MSD for both components in the
mixture. Differently from the single case, we observe a significant
slowing down also in the small component behavior, suggested by the
intermediate time regime where the slope of $\left<r_i^2\right>$
versus time (in the double-logarithmic plot) is smaller than $1$,
indicating a subdiffusive regime in the dynamics.  In the same time
window, the large stars are more confined than the small ones; their
localization length $l^\mathrm{MD}$ remains equal to that of the
single glass.  For the small stars we can deduce instead a
localization length of the order of $0.5\sigma_1$, about twice as
large as that of the large stars, again in agreement with MCT
predictions.

\begin{figure}
\centering
\includegraphics[width=8.5cm,clip]{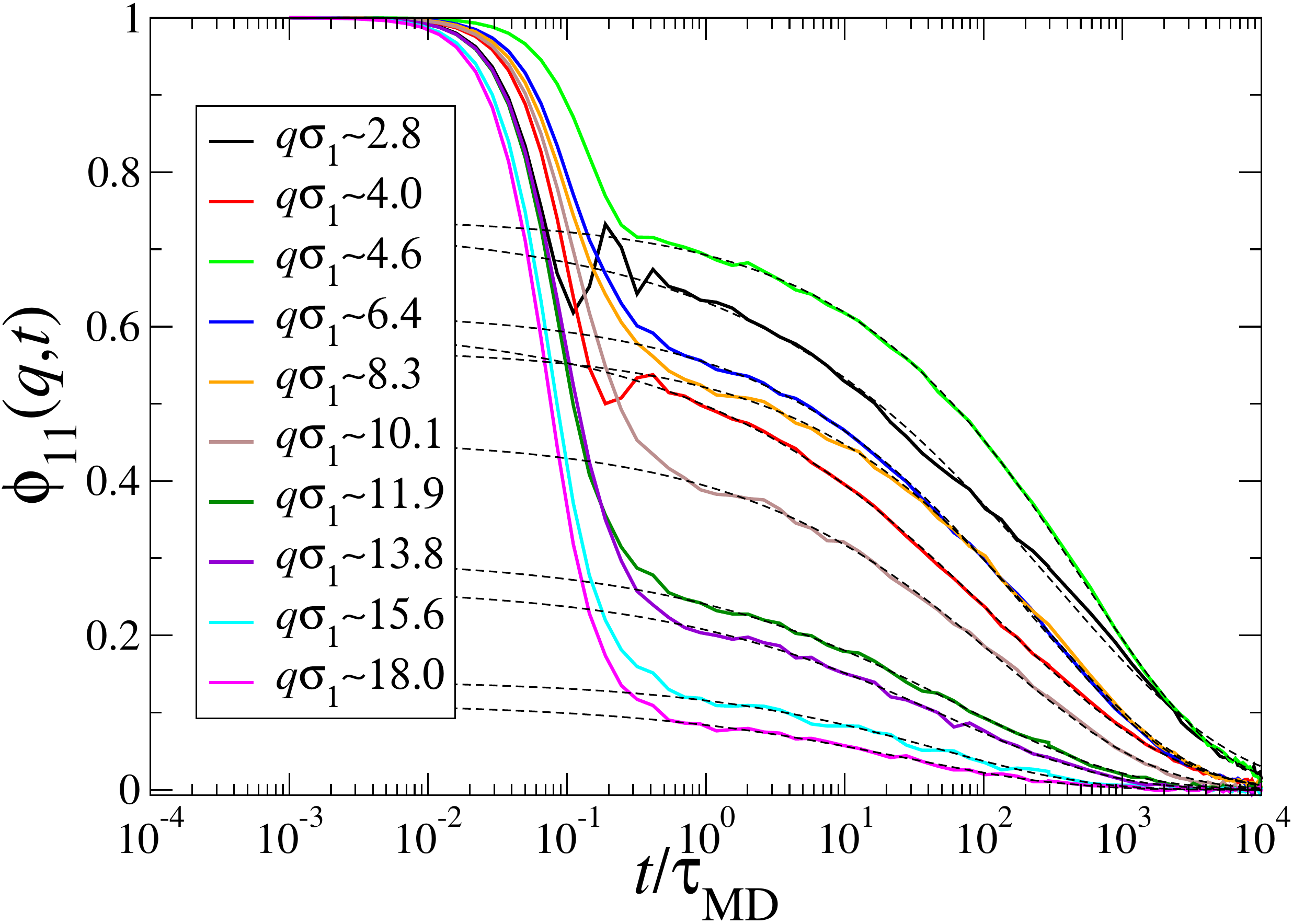}
\includegraphics[width=8.5cm,clip]{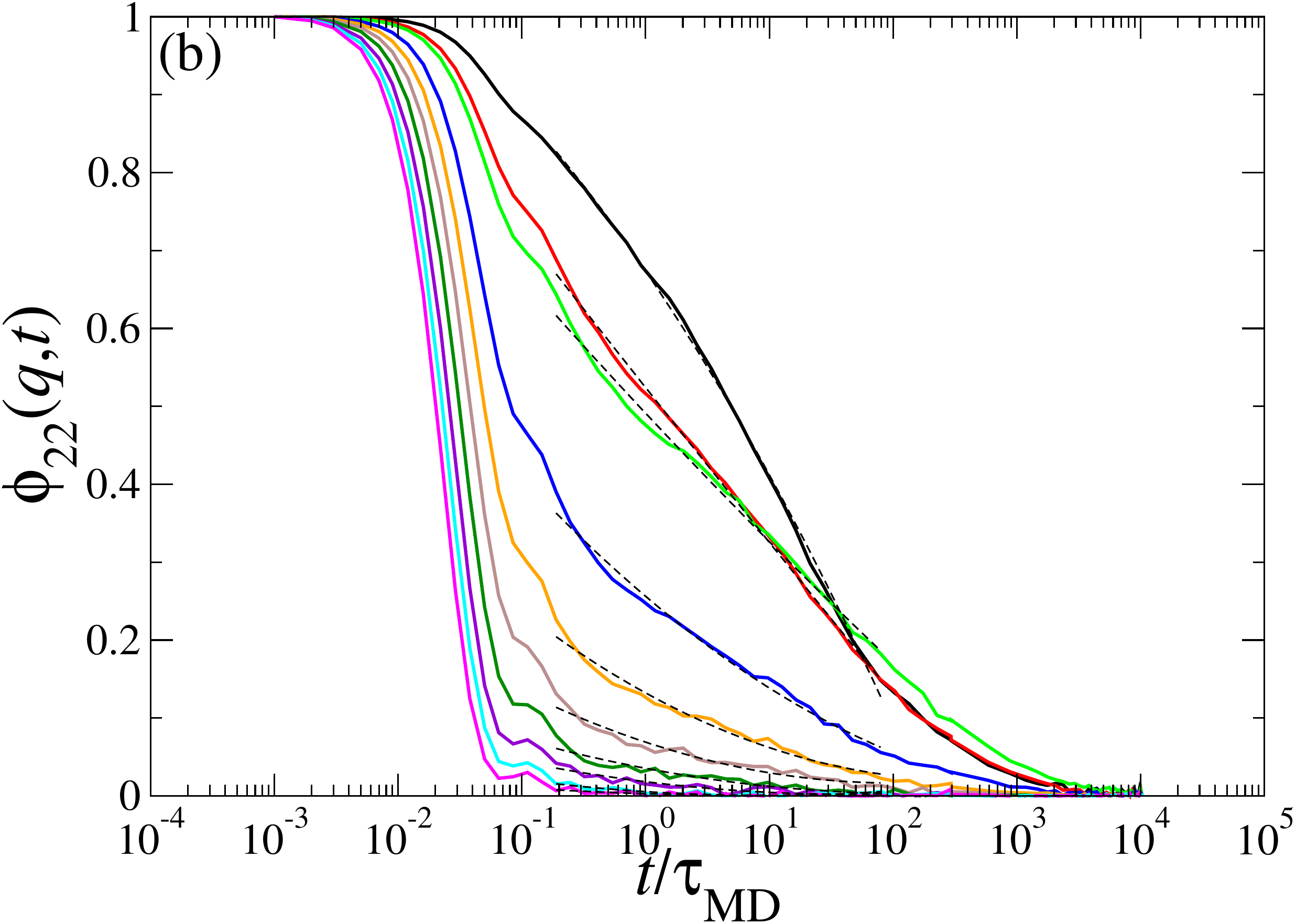}
\caption{Density correlators for large (a) and small (b) stars in the
  double glass for different values of wave number $q$ (see legend).
  The parameters are the same as in Fig.~\ref{fig:msd-DG}.  The dashed
  lines for the large stars are stretched exponential fits
  (Eq.\ (\ref{stretched})), while for the small stars we use
  logarithmic fits (Eq.\ (\ref{eq:mct-log})) with $\tau/\tau_{MD}=5$.}
\label{fig:sqt-DG}
\end{figure}

\begin{figure}
\centering
\includegraphics[width=8cm,clip]{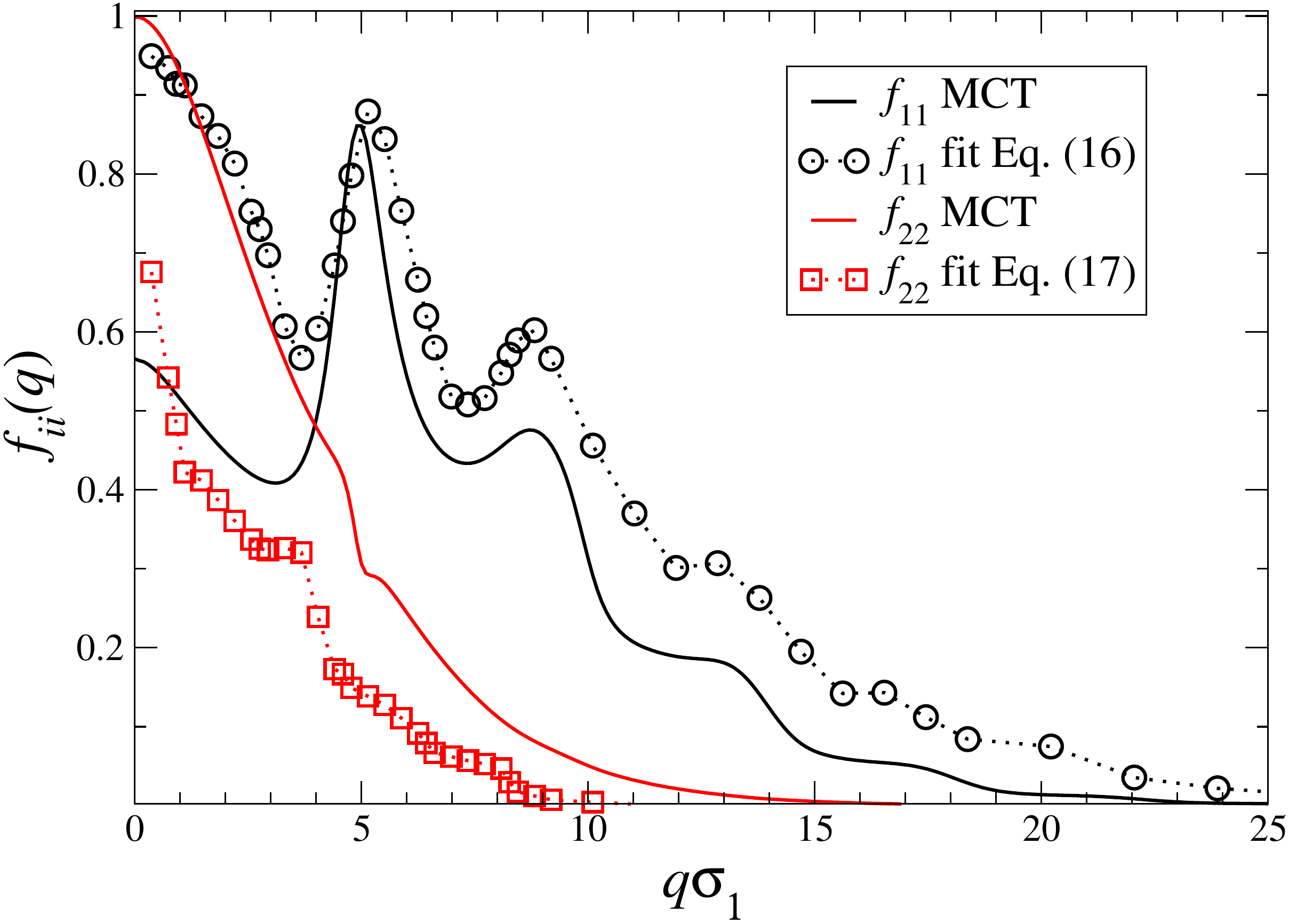}
\caption{
Comparison of partial non-ergodicity parameters from MCT and simulation in the double glass.
The values from the simulations are obtained by
stretched exponential (Eq.~(\ref{stretched})) fits for the large stars and from logarithmic (Eq.~(\ref{eq:mct-log})) fits for the small stars. 
The simulation parameters are the same as in Fig.~\ref{fig:msd-DG}.
The MCT results refer to the critical parameters: $\rho_1\sigma_1^3=0.345$, $\rho_2\sigma_1^3=0.05$ and $\delta=0.4$.
}
\label{fig:fq-DG}
\end{figure}

Looking at the partial density correlators for the double glass
(Fig.\ \ref{fig:sqt-DG}(a)), we see a close similarity for the large
stars correlators to those of the single glass.  Only the time
duration of the plateau changes, due to the different relative
position of the chosen state point with respect to the glass
transition.  Indeed, it can be seen in Figs.\ \ref{fig:mctphase} and
\ref{fig:iso} that this state point is `more liquid' than the one we
studied for the single glass.  Similarly to the previous paragraph, we
can analyze the decay in terms of a stretched exponential and of a von
Schweidler law and the results are totally equivalent to those of the
single glass concerning dynamics of the large stars. The resulting
$f_{11}(q)$ is reported in Fig.\ \ref{fig:fq-DG} together with
corresponding critical MCT predictions in this regime.

A completely different picture arises for the small stars
(Fig.\ \ref{fig:sqt-DG}(b)). The density correlators cannot be
described in terms of stretched exponentials. Indeed, they are best
fitted with second order polynomial in $\ln(t)$ of
Eq.\ (\ref{eq:mct-log}) with $\tau/\tau_{MD}=5$.  In other systems,
this anomalous dynamical behavior has been attributed to the
competition between two glassy states
\cite{Scio03,Zac06a,Moreno06JCP,Moreno06PRE,Moreno06condmat}. This
seems to suggest the possible existence of a higher order MCT
singularity \cite{gotzesperl}, although we do not detect an increase of the exponent parameter 
$\lambda$ in the MCT calculations. Interestingly, the logarithmic behavior
is found only for the small stars, while the behavior of the large
stars remains close to a standard liquid-glass transition or $A_2$
singularity.

Performing the fits following Eq.\ (\ref{eq:mct-log}), we extract a
non-ergodicity parameter (Fig.\ \ref{fig:fq-DG}) for the small
component in good agreement with the one calculated within MCT close
to the liquid-glass transition ($\rho_2\sigma_1^3=0.05$, see
Fig.\ \ref{fig:mctphase}).  Also, by looking at the $h_q^{(2)}$ (not
shown) extracted from the fits, we can find the wave vector $q^*$ at
which the decay is purely logarithmic \cite{gotzesperl,sperl04}, which
indicates the dominant length scale in the arrest process, and indeed
it was estimated before for the attractive glass \cite{Scio03}.  Here,
$h_q^{(2)}\sim 0 $ for $q\sigma_1\approx 4$, corresponding to a
length-scale slightly larger than the nearest-neighbor length of the
large stars, indicating that small stars are mostly trapped in the
voids between large stars. Around this $q^*$-value there is the
expected crossover between concave and convex shape for
$\phi_{22}(q,t)$ \cite{gotzesperl}.  As for the single glass case, the
estimates of $f_{ii}(q)$ from the simulations differ from the MCT ones
especially at low $q$, though the discrepancy is smaller in this case.

To summarize, the simulation results are in agreement with the MCT
predictions discussed above, which suggest that the glassy properties
of the large stars are identical in the single and double glass, both
in terms of non-ergodicity parameter and localization length, while
the small stars display a crossover from ergodic to arrested anomalous
behavior.

\subsection{Region III: Asymmetric Glass}

We analyze the state point $\rho_1=0.345$, $\rho_2\sigma_1^3=3.4$ and
$\delta=0.4$ (5378 particles in the simulation), where the diffusivity
decreases upon addition of small stars, as can be seen in
Fig.~\ref{fig:iso}.

\begin{figure}
\centering
\includegraphics[width=8cm,clip]{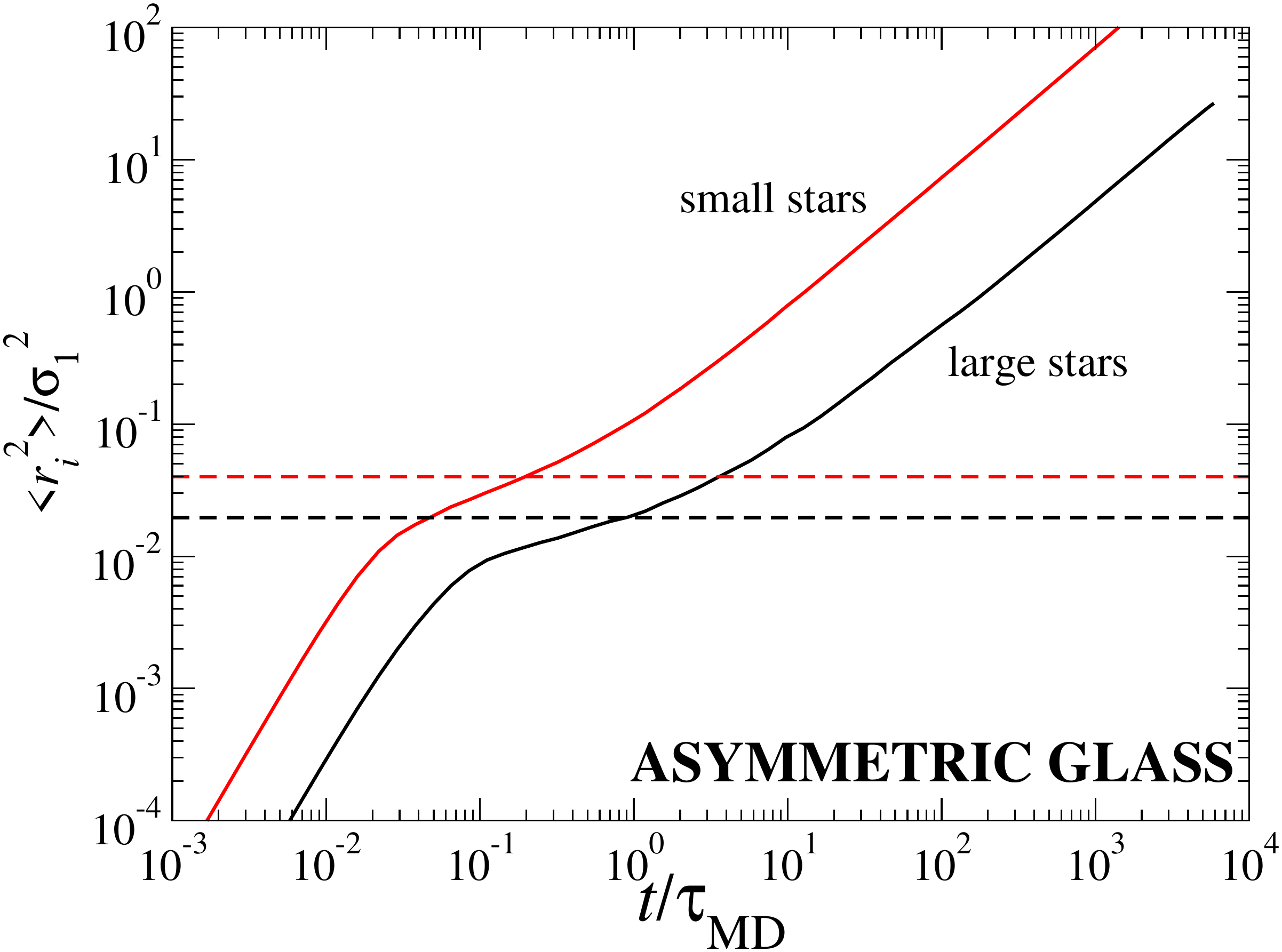}
\caption{Mean squared displacements for  large and small stars 
in the asymmetric glass. The densities are $\rho_1\sigma_1^3=0.345$
and   $\rho_1\sigma_1^3=3.4$; the size ratio is $\delta=0.4$. he horizontal lines indicate
the estimates of the squared localization length for  large and small stars, respectively, defined as the
inflection point of the MSD.}
\label{fig:msd-NG}
\end{figure}

From the MSD reported in Fig.~\ref{fig:msd-NG}, we see that both
components exhibit a slowing down in the dynamics, therefore we have
evidence of a second state in which both components become arrested,
i.e. a second double glass. However, the localization lengths that can
be extracted from the MSD are smaller than those for the double glass
discussed above. Indeed, we find $l_\mathrm{MD}\approx 0.14/\sigma_1$
for the large stars, corresponding to about half of the cage length
that we found both in single and double glass, while
$l_\mathrm{MD}\approx 0.2/\sigma_1$ for the small stars, again at
least twice as small as that of the other (conventional) double glass.
These values are again in remarkable agreement with those predicted by
MCT.  The additives do not only lead to a reduced localization length,
but also deform the cages of the big particles, leading to an
asymmetric glass \cite{unpublished}.

\begin{figure}
\centering
\includegraphics[width=8cm,clip]{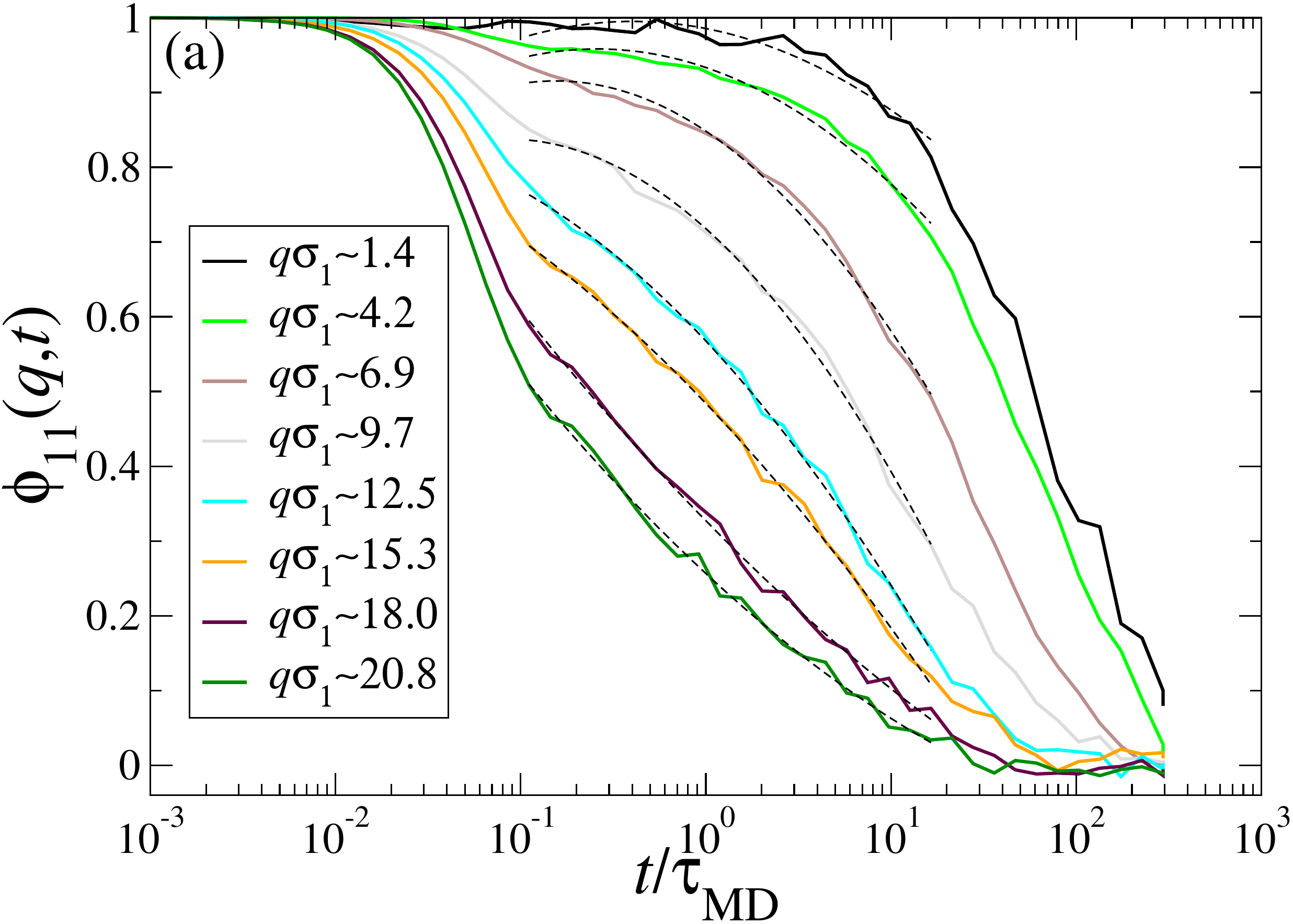}
\includegraphics[width=8cm,clip]{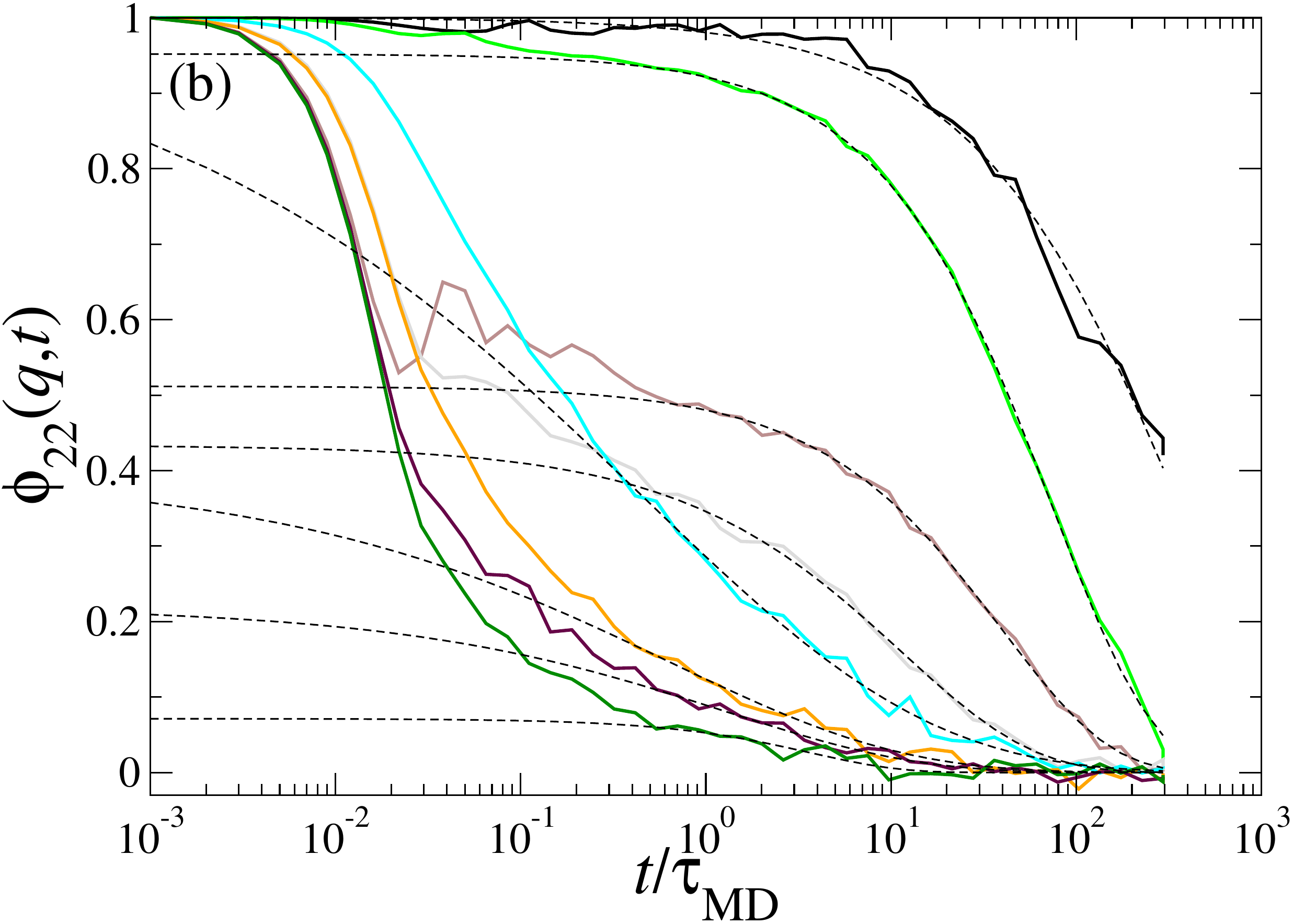}
\caption{Density correlators for  large (a) and small (b) stars
in the asymmetric for different wave numbers $q$ (see legend).
The parameters are the same as in Fig.~\ref{fig:msd-NG}.
Here the large stars are described by polynomial in $\ln(t)$  (Eq.~(\ref{eq:mct-log})), for the small ones we use stretched exponentials (Eq.~(\ref{stretched})).}
\label{fig:sqt-NG}
\end{figure}

The analysis of the density correlators provides evidence that the
dynamics in this double asymmetric glass is very different from that
of the standard double glass, see Fig.\ \ref{fig:sqt-NG}. Indeed, we
cannot fit the correlators of the large stars with stretched
exponentials, but rather they are found to follow the logarithmic
decay of Eq.\ (\ref{eq:mct-log}) with $\tau/\tau_{MD}=1$. On the other
hand, the small stars $\phi_{22}(q,t)$ are well described by a
stretched exponential decay for small wave vectors.  The logarithmic
behavior for large stars suggest the possibility of the presence of a
higher order MCT singularity for the large stars or, at least, a
competition between two different glassy states. Indeed, upon
increasing $\rho_2$ at fixed $\delta=0.4$, the large stars are passing
from the standard glass to a much more localized one.  The same is
true for the small stars in this region, so that a similar
$\log$-behavior is also expected for the small stars at some region of
the phase diagram, but within the present study we are not able to
detect that. However, we observe a deviation from standard stretched
exponential behavior at large $q$ also for the small stars.

\begin{figure}
\centering
\includegraphics[width=8cm,clip]{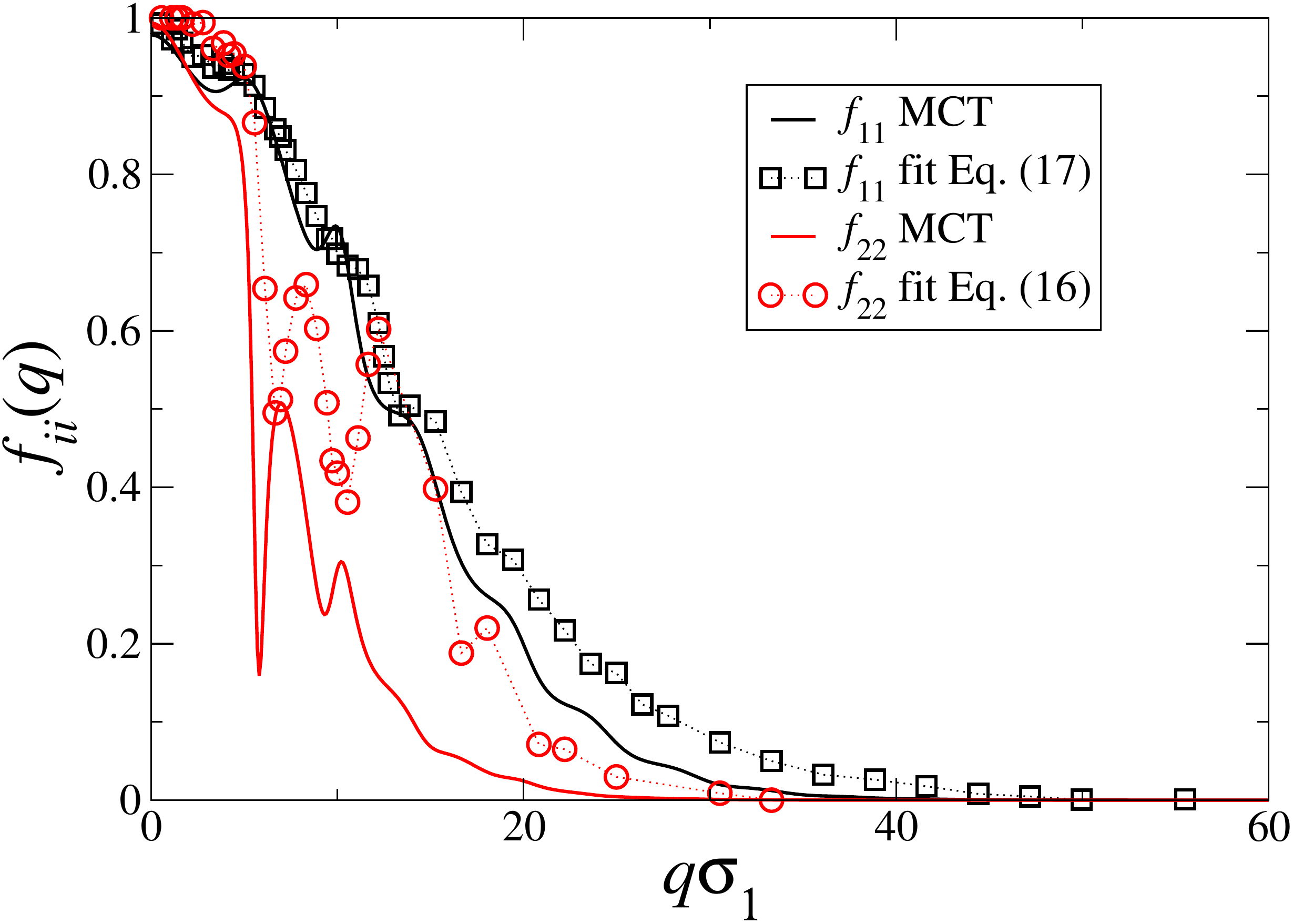}
\caption{ Comparison of partial non-ergodicity parameters from MCT and
  simulation in the asymmetric glass.  The values from the simulations
  are obtained by stretched exponential (Eq.~(\ref{stretched})) fits
  for the small stars and from logarithmic (Eq.~(\ref{eq:mct-log}))
  fits for the large stars with $\tau/\tau_{MD}=1$.  The simulation
  parameters are the same as in Fig.~\ref{fig:msd-NG}.  The MCT
  results refer to the critical parameters: $\rho_1\sigma_1^3=0.345$,
  $\rho_2\sigma_1^3=1.27$ and $\delta=0.4$.}
\label{fig:fq-NG}
\end{figure}

From the logarithmic fits for the large stars we determine the
non-ergodicity parameters for large stars, again in quite good
agreement with MCT predictions (see Fig.~\ref{fig:fq-NG}).  Also, we
can determine the $q^*$ where the time dependence is logarithmic,
i.e.\ $q^*\sigma_1\sim 16.6$, thus invoking smaller length scales than
the nearest-neighbor distance as responsible for the mechanism of
arrest.  We notice that this length is much larger than that found for
attractive glasses in CP mixtures \cite{Scio03}, consistent with the
fact that here no attractive bonding is present.  For the small stars,
the stretched exponential fits are good at small wave-vectors, but
become more uncertain for large $q$ values. However, the fits show a
satisfying agreement for the non-ergodicity parameters with the MCT
predictions for the critical parameters $\rho_1=0.345$, $\rho_2=1.27$
and $\delta=0.4$.  We have also tried to use a Von Schweidler exponent
$b= 0.5$, quite close to the MCT predictions, which also fits quite
well the small stars correlators.

To recapitulate, also in the asymmetric glass we find good agreement
between simulation results and MCT predictions. Moreover, we find
again anomalous dynamics. However, with respect to the double glass
the properties of the small and large stars density correlators are
reversed, i.e.\ now the large stars can be described in terms of
stretched exponentials while the small ones exhibit a logarithmic
decay.  This fact, together with the growth of $\lambda$ for this case
as reported in the MCT results section, points to the
possible existence of a nearby higher-order singularity for the large
stars.

\subsection{Region IV: Attractive Glass and Phase Separation}

Finally, we turn to the region of small $\delta$ and large $\rho_2$
where we expect the occurrence of the attractive glass.  However, this
is difficult to observe due to competition with phase separation in
the mixture. As already discussed before, we can only detect an
attractive glass with binary MCT, because we are not able to observe a
bending of the ideal glass line in the one-component MCT treatment
before the integral equation calculations cease to converge.

\begin{figure}[t]
\includegraphics[width=8cm, clip=true]{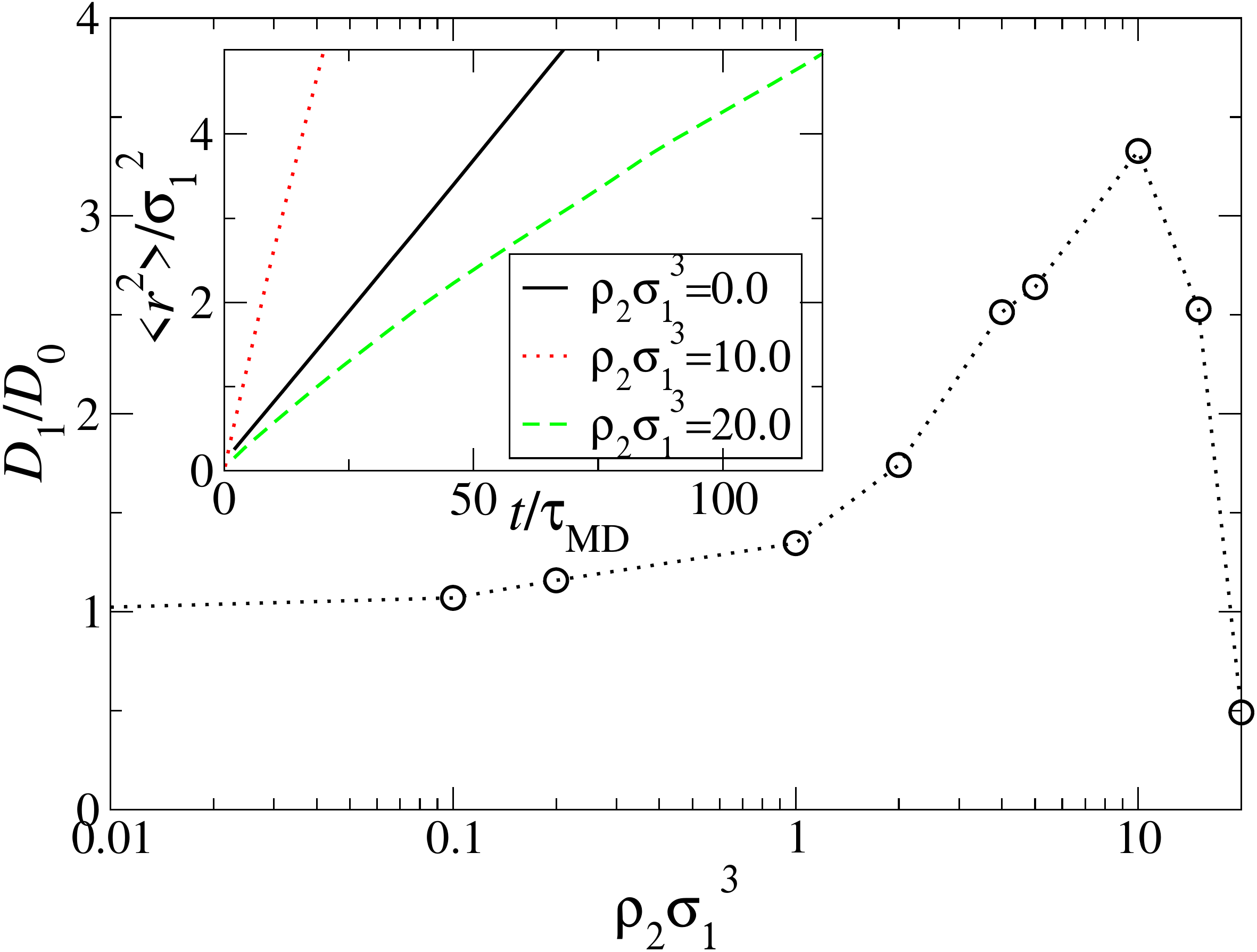}
\caption{Diffusion coefficient of the large stars, normalized by the value for the reference
one-component system $D_0$,  for $\rho_1\sigma_1^3=0.345$ and $\delta=0.1$
as a function of $\rho_2$. As an inset we show the MSD of the large stars
for different concentrations of small ones.}
\label{fig:diffattrglass}
\end{figure}

Nevertheless, in the simulations, we do observe a decrease of the
diffusion coefficient before the onset of phase separation
(Fig.\ \ref{fig:diffattrglass}).  After an initial increase of
diffusivity the dynamics becomes slower with increasing $\rho_2$.
However, this decrease is not sufficient to observe the emergence of a
plateau in the MSD and to extract values for the nonergodicity
parameter from the simulation.

\begin{figure}[t]
\includegraphics[width=8cm, clip=true]{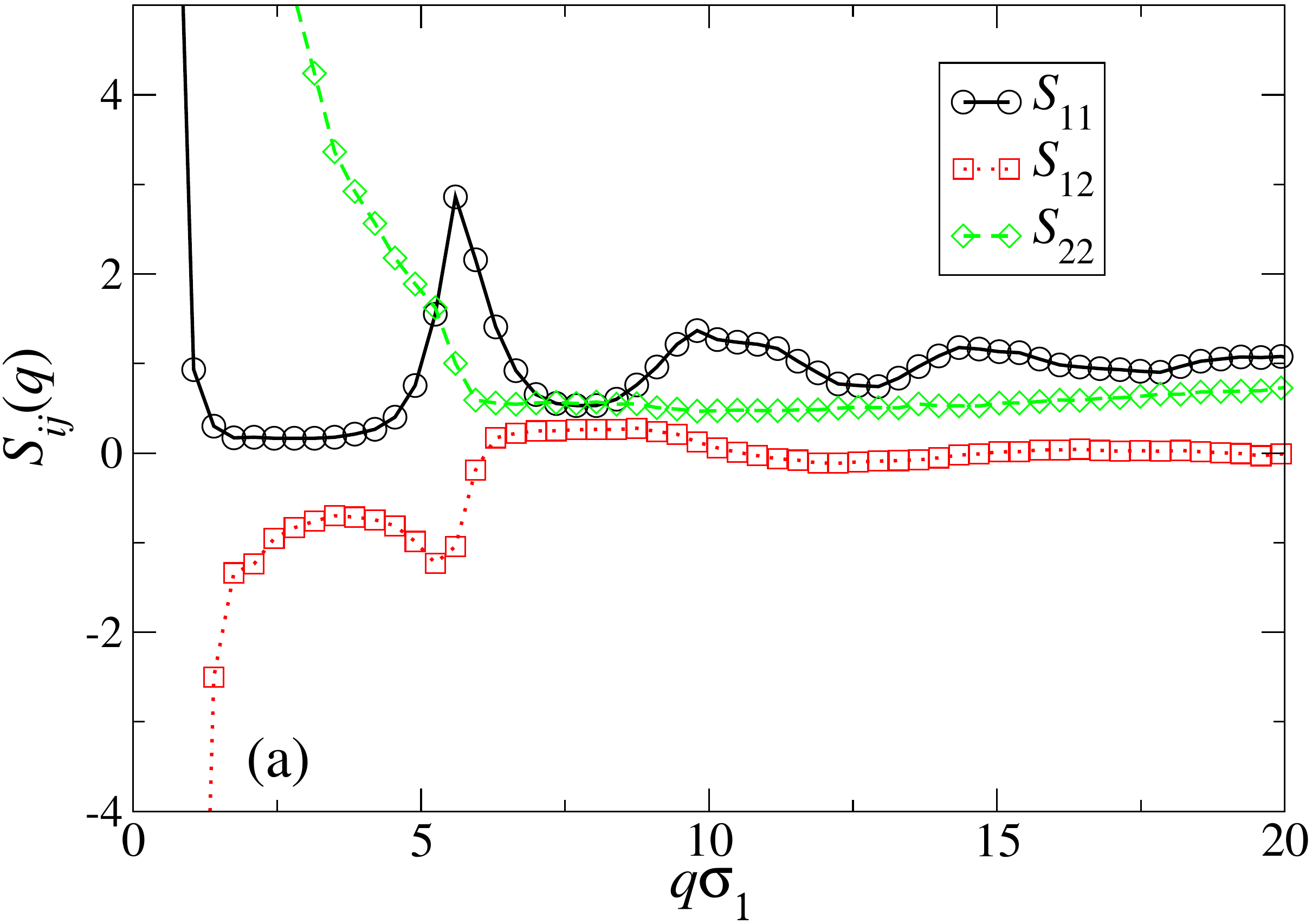}
\includegraphics[width=8cm, clip=true]{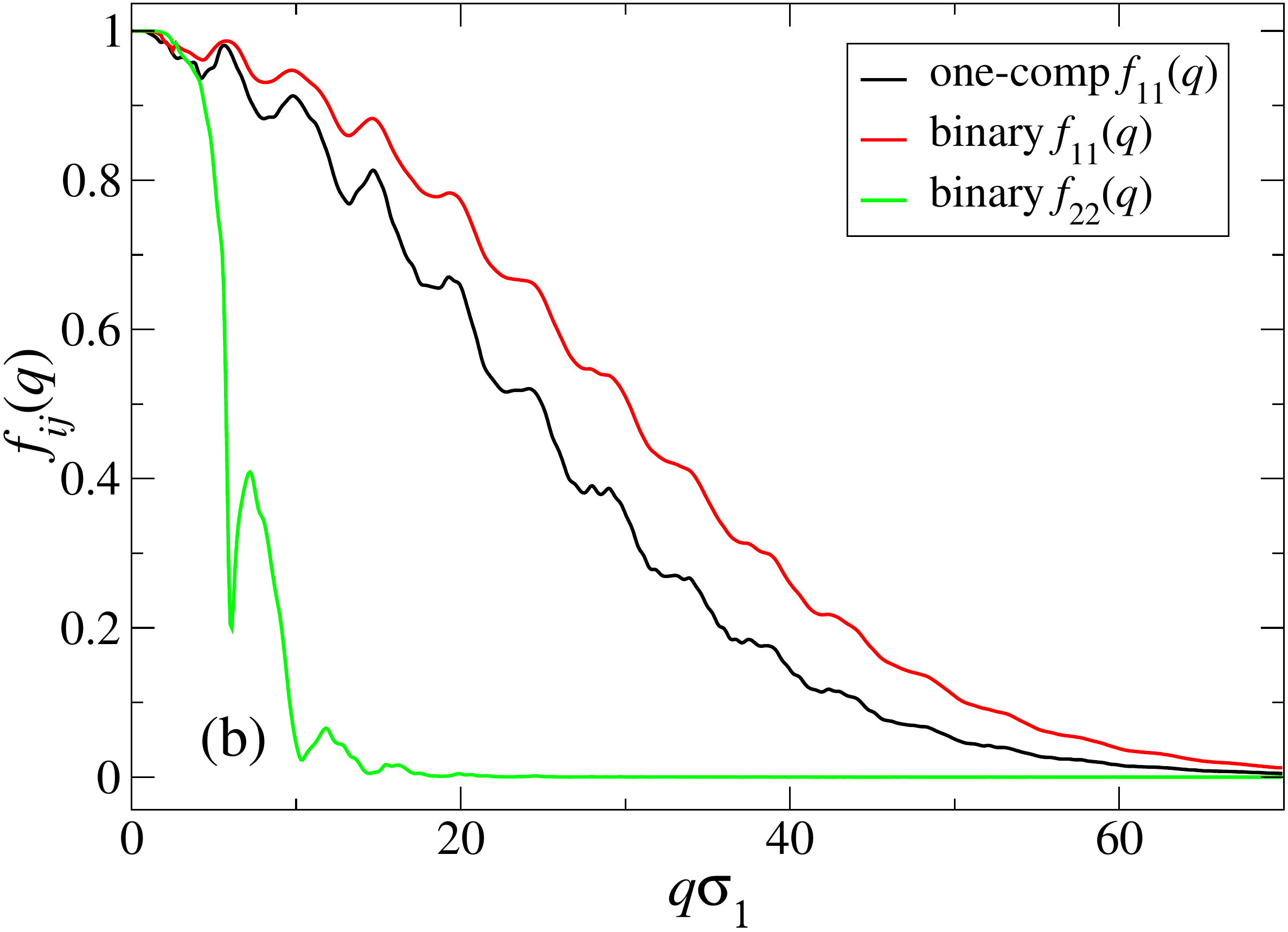}
\caption{(a) Partial structure factors $\rho_1\sigma_1^3=0.345$, $\delta=0.1$, and $\rho_2\sigma_1^3=20$. Note the
small-$q$ behavior of all three curves, which signals phase separation. (b) The corresponding non-ergodicity parameters
predicted by MCT.}
\label{fig:sqattrglass}
\end{figure}

We then used the static structure factor from the (monodisperse)
simulations as input for the mode-coupling equations and detected a
glassy state both in binary and one-component MCT for
$\rho_1\sigma_1^3=0.345$, $\delta=0.1$, and $\rho_2\sigma_1^3=20$.
The partial structure factors are shown in
Fig.\ \ref{fig:sqattrglass}(a). The small wave-length behavior of the
structure factors shows the typical increase of a state point close to
phase separation.  In Fig.\ \ref{fig:sqattrglass}(b) we show the
corresponding non-ergodicity parameters whose shape looks similar to
the one found in previous studies of systems with attractive
interactions \cite{ber99a}.  It is interesting to note that the one-
and two-component MCT give almost the same results for $f_{11}(q)$.
Hence, the decrease of the diffusivity combined with large-scale
fluctuations seems to indicate the possibility of an ``arrested phase
separation''\cite{zacca_review} in this case, the interplay beween
glass transition and phase separation induces the dynamical arrest of
the denser phase. This mechanism has been recently proposed to
describe dynamical arrest at low and intermediate densities in CP
mixtures \cite{Lu} and other attractive systems \cite{cardinaux}.

\section{Conclusions}

In conclusion, we have identified different glassy states in a binary 
mixture of star polymers
 by using extensive computer simulations and mode coupling theory. 
In the plane spanned by size asymmetry between the two components
and concentration of the added small star,
there is a liquid lake which reveals multiple reentrant behaviors. Both the upper and  lower threshold 
concentrations
for the liquid-glass transition behave nonmonotonically with the size asymmetry 
$\delta$. At high
concentrations $\rho_2$ of small stars and $\delta=0.4$, we have 
 identified a new, asymmetric double glass type which is characterized by a strong 
localization of the small particles and anisotropic cages. This is different
both from the usual double glass at low concentrations $\rho_2$ and from the hard 
sphere double glass \cite{Voigtmann}. Recent experiments confirm our theoretical findings \cite{unpublished}.
At low $\delta$ and high $\rho_2$ there is a competition between 
vitrification and phase separation. In all the explored phase space, we find remarkable agreement between simulation
results and MCT predictions.

An interesting open question is
which minimum arm numbers $f_1$ and $f_2$ are necessary for
the formation of a high-$\rho_2$ glass, as we know that one-component star polymer systems
with less than $\approx 40-45$ arms do not ever glassify \cite{foffi03}.
An additional interesting question concerns the presence of higher order singularities 
in the wide parameter space that can be explored with star-polymer binary mixtures.
Indeed, the MD results provide evidence of a non-conventional slowing down of the 
dynamics in the proximity of the double and of the asymmetric glass. It is important to find out if 
 MCT predicts any sharp glass-glass
transition between single and double glass, or between double and asymmetric
glass. This will be the subject of a
future study.  A more exhaustive theoretical
study is requested to properly map the MCT results with the numerical ones and
provide a guide-line for locating, if existing, the higher order singularities, similarly to what has been done for the case of short-range attractive 
colloids\cite{Scio03,sperl04}.

\begin{acknowledgments}
The authors thank E.~Stiakakis, D.~Vlassopoulos, and I. Saika-Voivod
for many helpful discussions.
This work has been supported  by the DFG
within the SFB-TR6, 
by MIUR PRIN and by the EU within the Network of Excellence
``Softcomp'' and
the Marie Curie Research and Training Network
``Dynamical Arrested States of Soft Matter and Colloids'' 
(MRTN-CT-2003-504712). 
C.M.\ thanks the Alexander von Humboldt Foundation for financial support.
C.N.L.\ thanks the Erwin Schr{\"o}dinger Institute (ESI)
in Vienna for 
a Senior Research Fellowship as well as the ESI and 
the Vienna University of Technology,
where parts of this work have 
been carried out, for their hospitality.
\end{acknowledgments}




\end{document}